\pdfoutput=1

\documentclass[12pt,a4paper]{article}

\usepackage{ifthen}
\newboolean{pdflatex}
\setboolean{pdflatex}{true}

\newboolean{articletitles}
\setboolean{articletitles}{true}

\newboolean{uprightparticles}
\setboolean{uprightparticles}{false}

\def\paperauthors{LHCb collaboration}
\def\paperasciititle{Measurement of the CP-violating phase phi_s from Bs->J/psi pi+pi- decays in 13 TeV pp collisions}
\def\papertitle{Measurement of the \CP-violating phase $\phi_s$ from $\Bs\to\jpsi\pi^+\pi^-$ decays in 13\tev $pp$ collisions}
\def\paperkeywords{{High Energy Physics}, {LHCb}} 
\def\papercopyright{\the\year\ CERN for the benefit of the LHCb collaboration} 
\def\paperlicence{CC-BY-4.0 licence}
\def\paperlicenceurl{https://creativecommons.org/licenses/by/4.0/}

\def\BsorBsbar {\kern \thebaroffset\optbar{\kern -\thebaroffset \Bs}\xspace}
\usepackage[top=1in, bottom=1.25in, left=1in, right=1in]{geometry}

\usepackage{microtype}
\usepackage{lineno}
\usepackage{xspace}
\usepackage{caption}

\usepackage{graphicx}
\usepackage{color}
\usepackage{colortbl}
\DeclareGraphicsExtensions{.pdf,.PDF,png,.PNG}

\usepackage{amsmath} 
\usepackage{amssymb}
\usepackage{amsfonts}
\usepackage{upgreek}

\newcommand*\patchAmsMathEnvironmentForLineno[1]{%
\expandafter\let\csname old#1\expandafter\endcsname\csname #1\endcsname
\expandafter\let\csname oldend#1\expandafter\endcsname\csname
end#1\endcsname
 \renewenvironment{#1}%
   {\linenomath\csname old#1\endcsname}%
   {\csname oldend#1\endcsname\endlinenomath}%
}
\newcommand*\patchBothAmsMathEnvironmentsForLineno[1]{%
  \patchAmsMathEnvironmentForLineno{#1}%
  \patchAmsMathEnvironmentForLineno{#1*}%
}
\AtBeginDocument{%
\patchBothAmsMathEnvironmentsForLineno{equation}%
\patchBothAmsMathEnvironmentsForLineno{align}%
\patchBothAmsMathEnvironmentsForLineno{flalign}%
\patchBothAmsMathEnvironmentsForLineno{alignat}%
\patchBothAmsMathEnvironmentsForLineno{gather}%
\patchBothAmsMathEnvironmentsForLineno{multline}%
\patchBothAmsMathEnvironmentsForLineno{eqnarray}%
}

\usepackage{hyperxmp}

\usepackage[pdftex,
            pdfauthor={\paperauthors},
            pdftitle={\paperasciititle},
            pdfkeywords={\paperkeywords},
            pdfcopyright={Copyright (C) \papercopyright},
            pdflicenseurl={\paperlicenceurl}]{hyperref}

\usepackage[colorinlistoftodos,textsize=scriptsize]{todonotes}

\usepackage[all]{hypcap} 

\def\lhcb   {\mbox{LHCb}\xspace}

\def\MagUp {\mbox{\em Mag\kern -0.05em Up}\xspace}

\ifthenelse{\boolean{uprightparticles}}%
{

 \def\Ppi         {\ensuremath{\uppi}\xspace}

 \def\Ppsi        {\ensuremath{\uppsi}\xspace}

 \def\PDelta      {\ensuremath{\Delta}\xspace}                 
 \def\PXi         {\ensuremath{\Xi}\xspace}                 
 \def\PLambda     {\ensuremath{\Lambda}\xspace}                 
 \def\PSigma      {\ensuremath{\Sigma}\xspace}                 
 \def\POmega      {\ensuremath{\Omega}\xspace}                 
 \def\PUpsilon    {\ensuremath{\Upsilon}\xspace}

 \def\PB      {\ensuremath{\mathrm{B}}\xspace}                 
                  
 \def\PD      {\ensuremath{\mathrm{D}}\xspace}

 \def\PJ      {\ensuremath{\mathrm{J}}\xspace}                 
 \def\PK      {\ensuremath{\mathrm{K}}\xspace}

 \def\Pb      {\ensuremath{\mathrm{b}}\xspace}                 
 \def\Pc      {\ensuremath{\mathrm{c}}\xspace}

 \def\Pi      {\ensuremath{\mathrm{i}}\xspace}

 \def\Ps      {\ensuremath{\mathrm{s}}\xspace}

 \def\thebaroffset{0.0em}
}
{

 \def\Ppi         {\ensuremath{\pi}\xspace}

 \def\Ppsi        {\ensuremath{\psi}\xspace}                 
                  
 \mathchardef\PDelta="7101
 \mathchardef\PXi="7104
 \mathchardef\PLambda="7103
 \mathchardef\PSigma="7106
 \mathchardef\POmega="710A
 \mathchardef\PUpsilon="7107
                  
 \def\PB      {\ensuremath{B}\xspace}                 
                  
 \def\PD      {\ensuremath{D}\xspace}

 \def\PJ      {\ensuremath{J}\xspace}                 
 \def\PK      {\ensuremath{K}\xspace}

 \def\Pb      {\ensuremath{b}\xspace}                 
 \def\Pc      {\ensuremath{c}\xspace}

 \def\Pi      {\ensuremath{i}\xspace}

 \def\Ps      {\ensuremath{s}\xspace}

 \def\thebaroffset{0.18em}
}
\newcommand{\offsetoverline}[2][\thebaroffset]{\kern #1\overline{\kern -#1 #2}}%

\makeatletter
\ifcase \@ptsize \relax%
  \newcommand{\miniscule}{\@setfontsize\miniscule{4}{5}}%
\or%
  \newcommand{\miniscule}{\@setfontsize\miniscule{5}{6}}%
\or%
  \newcommand{\miniscule}{\@setfontsize\miniscule{5}{6}}%
\fi
\makeatother

\DeclareRobustCommand{\optbar}[1]{\shortstack{{\miniscule (\rule[.5ex]{1.25em}{.18mm})}
  \\ [-.7ex] $#1$}}

\def\squark    {{\ensuremath{\Ps}}\xspace}

\def\cquark    {{\ensuremath{\Pc}}\xspace}

\def\bquark    {{\ensuremath{\Pb}}\xspace}

\def\pion   {{\ensuremath{\Ppi}}\xspace}

\def\pip    {{\ensuremath{\pion^+}}\xspace}
\def\pim    {{\ensuremath{\pion^-}}\xspace}
\def\pipm   {{\ensuremath{\pion^\pm}}\xspace}

\def\kaon    {{\ensuremath{\PK}}\xspace}

\def\KorKbar {\kern \thebaroffset\optbar{\kern -\thebaroffset \PK}{}\xspace}

\def\Kp      {{\ensuremath{\kaon^+}}\xspace}
\def\Km      {{\ensuremath{\kaon^-}}\xspace}

\def\D       {{\ensuremath{\PD}}\xspace}

\def\DorDbar {\kern \thebaroffset\optbar{\kern -\thebaroffset \PD}\xspace}

\def\Dsm     {{\ensuremath{\D^-_\squark}}\xspace}

\def\B       {{\ensuremath{\PB}}\xspace}
\def\Bbar    {{\ensuremath{\offsetoverline{\PB}}}\xspace}

\def\BorBbar {\kern \thebaroffset\optbar{\kern -\thebaroffset \PB}\xspace}
\def\Bz      {{\ensuremath{\B^0}}\xspace}

\def\Bu      {{\ensuremath{\B^+}}\xspace}

\def\Bp      {{\ensuremath{\Bu}}\xspace}

\def\Bd      {{\ensuremath{\B^0}}\xspace}
\def\Bs      {{\ensuremath{\B^0_\squark}}\xspace}
\def\Bsb     {{\ensuremath{\Bbar{}^0_\squark}}\xspace}

\def\Bc      {{\ensuremath{\B_\cquark^+}}\xspace}

\def\BorBsbar {\kern \thebaroffset\optbar{\kern -\thebaroffset \PB_s}\xspace}

\def\jpsi     {{\ensuremath{{\PJ\mskip -3mu/\mskip -2mu\Ppsi\mskip 2mu}}}\xspace}

\def\Y#1S{\ensuremath{\PUpsilon{(#1S)}}\xspace}

\def\Lz          {{\ensuremath{\PLambda}}\xspace}

\def\LorLbar     {\kern \thebaroffset\optbar{\kern -\thebaroffset \PLambda}\xspace}

\def\Lb           {{\ensuremath{\Lz^0_\bquark}}\xspace}

\def\to                 {\ensuremath{\rightarrow}\xspace}

\def\CP                {{\ensuremath{C\!P}}\xspace}

\newcommand{\dms}{{\ensuremath{\Delta m_{\squark}}}\xspace}

\newcommand{\DGs}{{\ensuremath{\Delta\Gamma_{\squark}}}\xspace}

\newcommand{\Gs}{{\ensuremath{\Gamma_{\squark}}}\xspace}

\newcommand{\GL}{{\ensuremath{\Gamma_{\mathrm{ L}}}}\xspace}
\newcommand{\GH}{{\ensuremath{\Gamma_{\mathrm{ H}}}}\xspace}

\newcommand{\phis}{{\ensuremath{\phi_{\squark}}}\xspace}

\def\AT#1     {\ensuremath{A_{\mathrm{T}}^{#1}}\xspace}           

\def\C#1      {\ensuremath{\mathcal{C}_{#1}}\xspace}                       
\def\Cp#1     {\ensuremath{\mathcal{C}_{#1}^{'}}\xspace}                    
\def\Ceff#1   {\ensuremath{\mathcal{C}_{#1}^{\mathrm{(eff)}}}\xspace}        
\def\Cpeff#1  {\ensuremath{\mathcal{C}_{#1}^{'\mathrm{(eff)}}}\xspace}       
\def\Ope#1    {\ensuremath{\mathcal{O}_{#1}}\xspace}                       
\def\Opep#1   {\ensuremath{\mathcal{O}_{#1}^{'}}\xspace}                    

\newcommand{\braket}[2]{\ensuremath{\langle #1|#2\rangle}} %

\newcommand{\nospaceunit}[1]{\ensuremath{\text{#1}}}       
\newcommand{\aunit}[1]{\ensuremath{\text{\,#1}}}       

\newcommand{\tev}{\aunit{Te\kern -0.1em V}\xspace}
\newcommand{\gev}{\aunit{Ge\kern -0.1em V}\xspace}
\newcommand{\mev}{\aunit{Me\kern -0.1em V}\xspace}
\newcommand{\kev}{\aunit{ke\kern -0.1em V}\xspace}
\newcommand{\ev}{\aunit{e\kern -0.1em V}\xspace}
\newcommand{\mevc}{\ensuremath{\aunit{Me\kern -0.1em V\!/}c}\xspace}
\newcommand{\gevc}{\ensuremath{\aunit{Ge\kern -0.1em V\!/}c}\xspace}
\newcommand{\mevcc}{\ensuremath{\aunit{Me\kern -0.1em V\!/}c^2}\xspace}
\newcommand{\gevcc}{\ensuremath{\aunit{Ge\kern -0.1em V\!/}c^2}\xspace}

\def\m    {\aunit{m}\xspace}

\def\mum  {\ensuremath{\,\upmu\nospaceunit{m}}\xspace}

\def\ps   {\ensuremath{\aunit{ps}}\xspace}
\def\fs   {\aunit{fs}}

\def\invps{\ensuremath{\ps^{-1}}\xspace}

\newcommand{\chisqip}{\ensuremath{\chi^2_{\text{IP}}}\xspace}

\def\gsim{{~\raise.15em\hbox{$>$}\kern-.85em
          \lower.35em\hbox{$\sim$}~}\xspace}
\def\lsim{{~\raise.15em\hbox{$<$}\kern-.85em
          \lower.35em\hbox{$\sim$}~}\xspace}

\newcommand{\Real}{\ensuremath{\mathcal{R}e}\xspace}
\newcommand{\Imag}{\ensuremath{\mathcal{I}m}\xspace}

\def\sPlot{\mbox{\em sPlot}\xspace}

\def\pt         {\ensuremath{p_{\mathrm{T}}}\xspace}

\def\ptot       {\ensuremath{p}\xspace}

\def\evtgen     {\mbox{\textsc{EvtGen}}\xspace}

\def\geant      {\mbox{\textsc{Geant4}}\xspace}

\def\photos     {\mbox{\textsc{Photos}}\xspace}

\def\pythia     {\mbox{\textsc{Pythia}}\xspace}

\xspace

\def\tell1  {TELL1\xspace}
\def\ukl1   {UKL1\xspace}

\newcommand{\mygevc}{\ensuremath{{\mathrm{Ge\kern -0.1em V\!/}c}}\xspace}
\newcommand{\xx}{\ensuremath{\kern 0.5em }}

\def\BorBbar    {\kern 0.18em\optbar{\kern -0.18em B}{}\xspace}

\newcommand{\Bsjpsipipi}{\ensuremath{\Bs \to\jpsi \pip \pim}\xspace}

\def\sPlot{\mbox{\em sPlot}\xspace}

\usepackage{multirow} %
\usepackage{booktabs} %
\usepackage{rotating}

\def \m {m_{\pi\pi}}
\def \angmu {\theta_{\jpsi}}
\def \angpi {\theta_{\pi\pi}}

\def \Bq {B_s^{0}}
\def \Bqb {\overline{B}{}_s^{0}}
\def \dv {{\rm d}}

\def \ch {\cosh \frac{\DGs t}{2}}
\def \sh {\sinh \frac{\DGs t}{2}}
\def \Ab {\overline{A}}
\def \cs {\cos(\dms t)}
\def \sn {\sin(\dms t)}

\def \A {{\cal A}}
\def \cAb {\overline{{\cal A}}}

\def \Bsdecay {$\Bs$ and $\Bsb\to \jpsi \pi^+ \pi^-$}
\DeclareRobustCommand{\optbar}[1]{\shortstack{{\miniscule (\rule[.5ex]{1.25em}{.18mm})}
  \\ [-.7ex] $#1$}}

\usepackage{epsfig,accents}
\newcommand*{\fancybar}{\scalebox{.4}{(}\raisebox{-1.7pt}{--}\scalebox{.4}{)}}
\newcommand*{\brabar}[1]{\accentset{\fancybar}{#1}}
\def\Bsorbar {{\ensuremath{{\kern \thebaroffset\optbar{\kern -\thebaroffset \PB}^0_{\!\!s}}}\xspace}}
\def\Gammaorbar{\kern \thebaroffset\optbar{\kern -\thebaroffset \Gamma}\xspace}
\def\omegaorbar{\kern \thebaroffset\optbar{\kern -\thebaroffset \omega}\xspace}

\usepackage{cite}
\usepackage{mciteplus}

\begin{document}

\renewcommand{\thefootnote}{\fnsymbol{footnote}}
\setcounter{footnote}{1}

\begin{titlepage}
\pagenumbering{roman}

\vspace*{-1.5cm}
\centerline{\large EUROPEAN ORGANIZATION FOR NUCLEAR RESEARCH (CERN)}
\vspace*{1.5cm}
\noindent
\begin{tabular*}{\linewidth}{lc@{\extracolsep{\fill}}r@{\extracolsep{0pt}}}
\ifthenelse{\boolean{pdflatex}}
{\vspace*{-1.5cm}\mbox{\!\!\!\includegraphics[width=.14\textwidth]{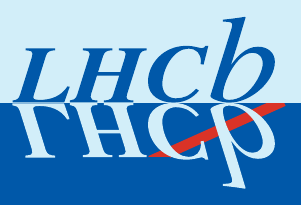}} & &}%
{\vspace*{-1.2cm}\mbox{\!\!\!\includegraphics[width=.12\textwidth]{lhcb-logo.eps}} & &}%
\\
 & & CERN-EP-2019-037 \\ 
 & & LHCb-PAPER-2019-003\\
 & &6 August 2019 \\ 
 & & \\
\end{tabular*}

\vspace*{2.0cm}

{\normalfont\bfseries\boldmath\huge
\begin{center}
  \papertitle 
\end{center}
}

\vspace*{1.5cm}

\begin{center}
\paperauthors\footnote{Authors are listed at the end of this paper.}
\end{center}

\vspace{1.0cm}

\begin{abstract}
  \noindent
  Decays of \Bs and \Bsb mesons into $\jpsi\pi^+\pi^-$ final states are studied in a data sample corresponding to 1.9~fb$^{-1}$ of integrated luminosity collected with the LHCb detector in 13~TeV $pp$ collisions. A time-dependent amplitude analysis is used to determine the final-state resonance contributions, the \CP-violating phase \mbox{$\phi_s=-0.057\pm 0.060\pm 0.011$~rad,} the decay-width difference between the heavier mass \Bs eigenstate and the $\Bz$ meson of $-0.050\pm 0.004\pm 0.004$~ps$^{-1}$, and the \CP-violating parameter \mbox{$|\lambda|=1.01_{-0.06}^{+0.08}\pm0.03$,} where the first uncertainty is statistical and the second systematic.  These results are combined with previous LHCb measurements in the same decay channel using 7~TeV and 8~TeV $pp$ collisions obtaining $\phi_s =0.002\pm0.044\pm0.012$~rad, and $|\lambda| =0.949\pm0.036\pm0.019$.
\end{abstract}

\vspace*{1.0cm}

\begin{center}
Published in Phys. Lett. B797 (2019)
  \end{center}

\vspace{\fill}

{\footnotesize 
\centerline{\copyright~\papercopyright. \href{\paperlicenceurl}{\paperlicence}.}}
\vspace*{2mm}

\end{titlepage}

\newpage
\setcounter{page}{2}
\mbox{~}
\cleardoublepage

\renewcommand{\thefootnote}{\arabic{footnote}}
\setcounter{footnote}{0}

\pagestyle{plain}
\setcounter{page}{1}
\pagenumbering{arabic}

\section{Introduction}
\label{sec:Introduction}
Measurements of \CP violation in final states that can be populated both by direct decay and via mixing provide an excellent way of looking for physics beyond the Standard Model (SM) \cite{Bigi:2000yz}. As yet unobserved heavy bosons, light bosons with extremely small couplings, or fermions can be present virtually in quantum loops, and thus affect the relative \CP phase. Direct decays into non-flavour-specific final states can interfere with those that undergo $\Bs-\Bsb$ mixing prior to decay. This interference can result in \CP violation. In certain \Bs decays one \CP-violating phase that can be measured, called $\phi_s$, can be expressed in terms of Cabibbo-Kobayashi-Maskawa matrix elements as $-2{\arg}\left[{-V_{ts}V_{tb}^*}/{V_{cs}V_{cb}^*}\right]$. 
It is not predicted in the SM, but can be inferred with high precision from other experimental data giving a value of $-36.5_{-1.2}^{+1.3}$\,mrad~\cite{CKMfitter2015}. This number is consistent with previous measurements, which did not have enough sensitivity to determine a non-zero value \cite{Aaltonen:2007he,*Abazov:2008af,*CDF:2011af,*Abazov:2011ry,LHCb-PAPER-2014-019,LHCb-PAPER-2014-059,LHCb-PAPER-2017-008,Aad:2016tdj,*Khachatryan:2015nza}. In this paper we present the results of a new analysis of the $\Bs\to\jpsi\pi^+\pi^-$ decay using data from 13\tev $pp$ collisions  collected using the LHCb detector in 2015 and 2016.\footnote{In this paper mention of a particular final state implies use of the charge-conjugate state, except when dealing with \CP-violating processes.} The existence of this decay and its use in \CP-violation studies was suggested in Ref.~\cite{Stone:2008ak}.

\section{Detector and simulation}
\label{sec:Detector}

The \lhcb detector~\cite{Alves:2008zz,LHCb-DP-2014-002} is a single-arm forward
spectrometer covering the \mbox{pseudorapidity} range $2<\eta <5$,
designed for the study of particles containing \bquark or \cquark
quarks. The detector includes a high-precision tracking system
consisting of a silicon-strip vertex detector surrounding the $pp$
interaction region~\cite{LHCb-DP-2014-001}, a large-area silicon-strip detector located
upstream of a dipole magnet with a bending power of about
$4{\mathrm{\,Tm}}$, and three stations of silicon-strip detectors and straw
drift tubes  placed downstream of the magnet.
The tracking system provides a measurement of the momentum, \ptot, of charged particles with
a relative uncertainty that varies from 0.5\% at low momentum to 1.0\% at 200\gev.\footnote{We use natural units where $\hbar=c=1$.}
The minimum distance of a track to a primary vertex (PV), the impact parameter (IP),
is measured with a resolution of $(15+29/\pt)\mum$,
where \pt is the component of the momentum transverse to the beam, in\,\gev.
Different types of charged hadrons are distinguished using information
from two ring-imaging Cherenkov detectors~\cite{LHCb-DP-2012-003}.
Photons, electrons and hadrons are identified by a calorimeter system consisting of
scintillating-pad and preshower detectors, an electromagnetic
and a hadronic calorimeter. Muons are identified by a
system composed of alternating layers of iron and multiwire
proportional chambers~\cite{LHCb-DP-2012-002}.
The online event selection is performed by a trigger,
which consists of a hardware stage, based on information from the calorimeter and muon
systems, followed by a software stage, which applies a full event
reconstruction.

At the hardware trigger stage, events are required to have a muon with high \pt or a hadron, photon or electron with high transverse energy in the calorimeters. 
The software trigger is composed of two stages, the first of which performs a partial reconstruction and requires either a pair of well-reconstructed, oppositely charged muons having an invariant mass above $2.7\gev$, or a single well-reconstructed muon with $\pt>1\gev$ and have a large IP significance $\chisqip>7.4$. The latter is defined as the difference in the $\chi^{2}$ of the vertex fit for a given PV reconstructed with and without the considered particles. The second stage applies a full event reconstruction and for this analysis requires two opposite-sign muons to form a good-quality vertex that is well-separated from all of the PVs, and to have an invariant mass within $\pm120\mev$ of the known \jpsi mass~\cite{PDG2018}.

 Simulation is required to model the effects of the detector acceptance and the
  imposed selection requirements.
  In the simulation, $pp$ collisions are generated using
\pythia~\cite{Sjostrand:2006za,*Sjostrand:2007gs}
 with a specific \lhcb
configuration~\cite{LHCb-PROC-2010-056}.  Decays of unstable particles
are described by \evtgen~\cite{Lange:2001uf}, in which final-state
radiation is generated using \photos~\cite{Golonka:2005pn}. The
interaction of the generated particles with the detector, and its response,
are implemented using the \geant
toolkit~\cite{Allison:2006ve, *Agostinelli:2002hh} as described in
Ref.~\cite{LHCb-PROC-2011-006}.

\section{Decay amplitude}
The resonance structure in \Bsdecay~decays has been previously studied with a time-integrated amplitude analysis  using 7 and 8\tev $pp$ collisions~\cite{LHCb-PAPER-2013-069}. The final state was found to be compatible with being entirely \CP-odd, with the \CP-even state fraction below $2.3$\% at 95\% confidence level, which allows the determination of the decay width of the heavy $\Bs$ mass eigenstate, $\GH$. The possible presence of a \CP-even component is taken into account when determining $\phi_s$ \cite{Zhang:2012zk}.

The total decay amplitude for a \BsorBsbar meson at decay time equal to zero is assumed to be the sum over individual $\pip\pim$ resonant transversity amplitudes \cite{Dighe:1995pd}, and one nonresonant amplitude, with each transversity component labelled as ${A}_i$ ($\overline{A}_i$).  Because of the spin-1 $\jpsi$ meson  in the final state, the three possible polarizations of the $\jpsi$ generate longitudinal ($0$), parallel ($\parallel$) and perpendicular ($\perp$) transversity amplitudes. When the $\pip\pim$ pair forms a spin-0 state the final system only has a longitudinal component, and thus is a pure \CP eigenstate. The parameter $\lambda_i \equiv \frac{q}{p}\frac{\Ab_i}{A_i}$, relates \CP violation in the interference between mixing and decay associated with the polarization state $i$ for each resonance in the final state. Here the quantities $q$ and $p$ relate the mass and flavour eigenstates, $p\equiv \braket{\Bs}{B_{\rm L}}$, and $q\equiv \braket{\Bsb}{B_{\rm L}}$, where $\vert B_{\rm L} \rangle$ is the lighter mass eigenstate~\cite{Bigi:2000yz}. The total amplitudes ${\cal A}$ and $\cAb$ can be expressed as the sums of the individual $\BsorBsbar$ amplitudes,  ${\cal A}=\sum A_i$ and $\cAb =\sum \frac{q}{p} \Ab_i =\sum \lambda_i A_i= \sum \eta_i |\lambda_i| e^{-i\phi_s^i} A_i$, with $\eta_i$ being the \CP eigenvalue of the state.  For each transversity state $i$ there is a \CP-violating phase $\phi_s^i\equiv -\arg(\eta_i\lambda_i)$~\cite{Aaij:2013oba}. Assuming that  \CP violation in the decay is the same for all amplitudes, then $\lambda\equiv \eta_i\lambda_i$ and $\phi_s\equiv -\arg(\lambda)$.  Using $|p/q|=1$, the decay rates for $\Bs$ and $\Bsb$ into the $\jpsi \pip\pim$ final state are\footnote{The latest LHCb measurement determined $|p/q|^2=1.0039\pm0.0033$ \cite{LHCb-PAPER-2016-013}.}
\begin{eqnarray}\label{Eq-t}
\Gammaorbar(t) \propto
 e^{-\Gs t}\left\{\frac{|\A|^2+|\cAb|^2}{2}\ch  \pm \frac{|\A|^2-|\cAb|^2}{2}\cs\right.\quad\quad\nonumber\\
- \left.\Real(\A^*\cAb)\sh  \mp  \Imag(\A^*\cAb)\sn\right\},
\end{eqnarray}
where the -- sign before the $\cs$ term and $+$ sign before the $\sn$ term apply to $\overline{\Gamma}(t)$, $\DGs \equiv \GL-\GH$ is the decay-width difference between the light and the heavy mass eigenstates, $\dms \equiv m_{\rm H}-m_{\rm L}$ is the corresponding mass difference, and $\Gs \equiv (\GL+\GH)/2$ is the average $\Bs$ meson decay width~\cite{Nierste:2009wg}.  

For $\jpsi$ decays to $\mu^+\mu^-$ final states the $A_i$ amplitudes are themselves functions of four variables: the $\pip\pim$ invariant mass $\m$, and  three angular variables \mbox{$\Omega\equiv (\cos\angpi, \cos\angmu, \chi)$}, defined in the helicity basis. These angles are defined as $\angpi$ between the $\pi^+$ direction in the $\pip\pim$ rest frame with respect to the $\pip\pim$ direction in the $\Bs$ rest frame,  $\angmu$ between the $\mu^+$ direction in the $\jpsi$ rest frame with respect to the $\jpsi$ direction in the $\Bs$ rest frame,  
and  $\chi$ between the $\jpsi$ and $\pip\pim$ decay planes in the $\Bs$ rest frame \cite{Zhang:2012zk,Aaij:2013oba}.  (These definitions are the same for $\Bq$ and $\Bqb$, namely, using $\mu^+$ and $\pi^+$ to define the angles for both $\Bq$ and $\Bqb$ decays.)
The explicit forms of the $|{\cal A}(\m,\Omega)|^2$,  $|\cAb(\m,\Omega)|^2$, and $\A^*(\m,\Omega)\cAb(\m,\Omega)$ terms in Eq.~(\ref{Eq-t}) are given in Ref.~\cite{Zhang:2012zk}.

The analysis proceeds 
by performing an unbinned maximum-likelihood fit to the \pip\pim mass distribution, the decay time, and helicity angles of \Bs candidates identified as \Bs or \Bsb by a 
flavour-tagging algorithm \cite{Aaij:2012mu,*Aaij:2016psi,*Fazzini:2018dyq}.\footnote{We utilize the same likelihood construction that we used to determine $\phi_s$ and $|\lambda|$ in $\Bs\to\jpsi K^+ K^-$ decays with $K^+K^-$ above the $\phi(1020)$ mass region~\cite{LHCb-PAPER-2017-008}.} The fit provides the \CP-even and \CP-odd components, and since we include the initial flavour tag, the fit also determines the \CP-violating parameters $\phis$ and $|\lambda|$, and the decay width.
In order to proceed, we need to select a clean sample of \Bs decays, determine acceptance corrections, perform a calibration of the decay-time resolution in each event as a function of its uncertainty, and calibrate the flavour-tagging algorithm.

\section{Selection requirements}

The selection of  $\jpsi \pip\pim$ right-sign (RS), and wrong-sign (WS) $\jpsi\pipm\pipm$ final states, proceeds in two phases. Initially we impose loose requirements and subsequently use a multivariate analysis to further suppress the combinatorial background. 
In the first phase we require that the \jpsi decay tracks be identified as muons, have $\pt>500\mev$, and form a good vertex with vertex fit $\chi^{2}$ less than 16. The identified pions are required to have $\pt>250\mev$, not originate from any PV, and form a good vertex with the muons. 
The resulting \Bs candidate is assigned to the PV for which it has the smallest \chisqip. Furthermore, we require that the smallest \chisqip is not greater than 25.
 The \Bs candidate is required to have its momentum vector aligned with the vector connecting the PV to the \Bs decay vertex, and to have a decay time greater than 0.3\ps. 
Reconstructed tracks sharing the same hits are vetoed. 

In addition, background from $\Bp\to\jpsi K^+$ decays,\footnote{When discussing flavour-specific decays, mention of a particular mode implies the additional use of the charge-conjugate mode.} where the $K^+$ is misidentified as a $\pi^+$ and combined with a random $\pi^-$, is vetoed by assuming that each detected pion is a kaon, computing the $\jpsi \Kp$ mass, 
and removing those candidates that  are within $\pm36$\mev of the known \Bp mass~\cite{PDG2018}. Backgrounds from  $\Bz\to \jpsi K^+\pi^-$ or $\Bs\to\jpsi K^+K^-$ decays with misidentified kaons result in masses lower than the \Bs peak and thus do not need to be vetoed.

For the multivariate part of the selection, we use a Boosted Decision Tree, BDT~\cite{Breiman,AdaBoost}, with the uBoost algorithm~\cite{Stevens:2013dya}. The algorithm is optimized to not further bias acceptance on the variable $\cos\angpi$. The variables used to train the BDT are the difference between the muon and pion identifications for the muon identified with lower quality, the \pt of the $\Bs$ candidate, the sum of the \pt of the two pions, and the natural logarithms of: the \chisqip of each of the pions, the $\chi^2$ of the $\Bs$ vertex and decay tree fits \cite{Hulsbergen:2005pu},
and the \chisqip of the $\Bs$ candidate. In the fit, the \Bs momentum vector is constrained to point to the PV, the two muons are constrained to the \jpsi mass, and all four tracks are constrained to originate from the same vertex.

Implementing uBoost requires a training procedure. Data background in the $\jpsi\pi^+\pi^-$ mass interval between 200 to 250 MeV above the $\Bs$ mass and simulated signal are first used. Then, separate samples are used to test the BDT performance. We weight the training simulation samples to match the two-dimensional $\Bs$ $p$ and $\pt$ distributions, and smear the vertex fit $\chi^2$, to match the background-subtracted preselected data. Finally, the minimum requirement for BDT point is chosen to maximize signal significance, $S/\sqrt{S+B}$, where $S(B)$ is the expected signal (background) yields in a range corresponding to $\pm2.5$ times the mass resolution around the known $\Bs$ mass~\cite{PDG2018}.

\begin{figure}[tb]
\centering
\includegraphics[width=0.65\textwidth]{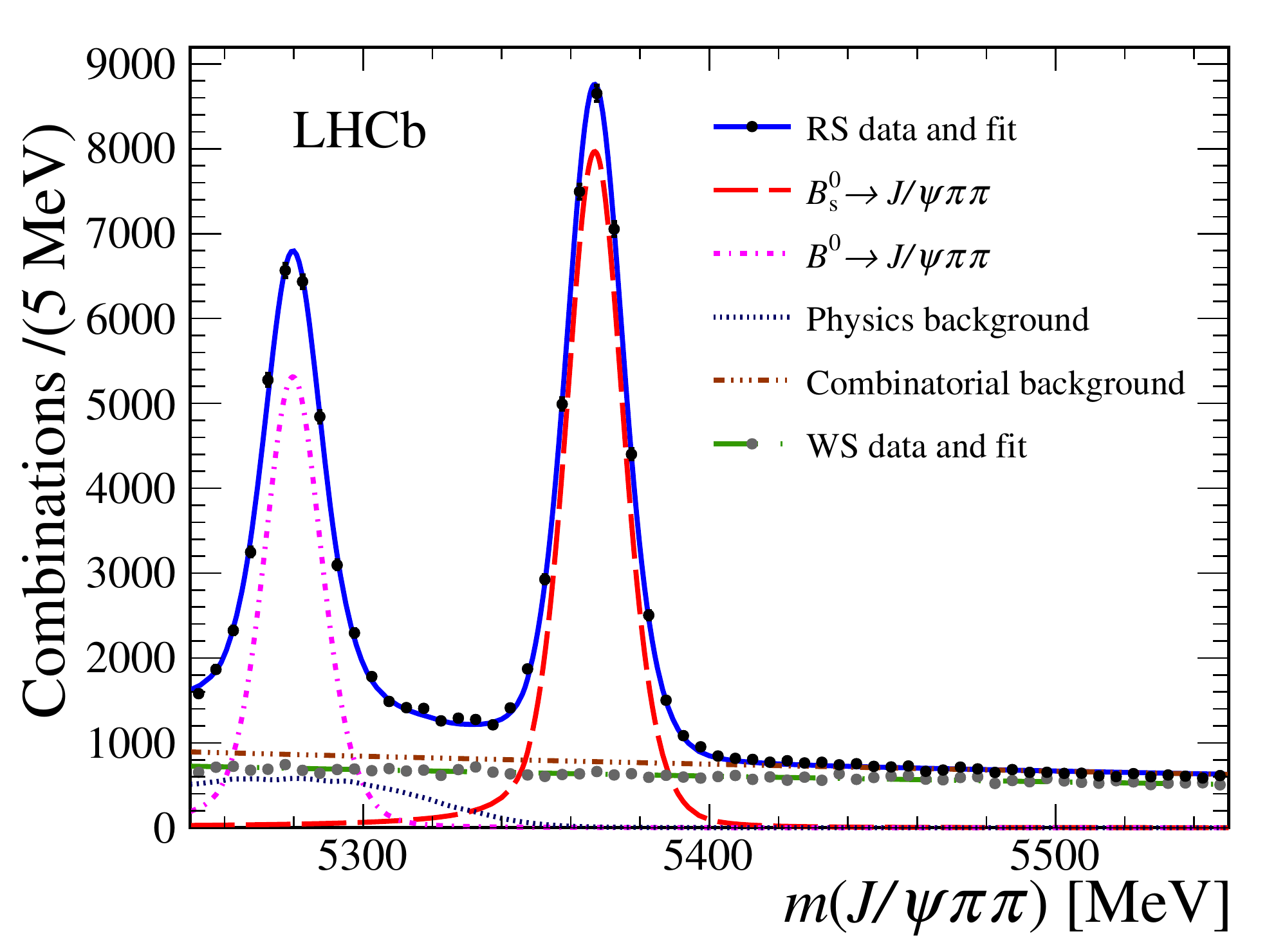}
\caption{Results of the simultaneous fit to the $\jpsi\pi\pi$ mass distributions RS (black points) and WS (grey points) samples. The solid (blue) curve shows the fit to the RS sample, the long dashed (red) curve shows the signal, the dot-dashed (magenta) curve shows $\Bd\to\jpsi \pip\pim$ decays,  the dot-long-dashed (brown) curve shows the combinatorial background, the dotted (black) curve shows the sum of $\Bs\to \jpsi \eta^\prime$ and $\Lb$ background, while the dot-dot-dashed (green) curve shows the fit to WS sample.}
\label{fig:massfit}
\end{figure}

To determine the signal and background yields we fit the candidate \Bs mass distribution.
Backgrounds include combinatorics, whose shape is estimated using WS $\jpsi\pi^{\pm}\pi^{\pm}$ candidates modelled by an exponential function, $\Bs\to\jpsi \eta^\prime (\to\rho^0 \gamma)$ decays with the $\gamma$ ignored, and $\Lb\to\jpsi p\Km$ decays with both hadrons misidentified as pions.  The latter backgrounds are modelled using simulation. The $\Bs$ signal shape is parameterized by a Hypatia function~\cite{Santos:2013gra}, where the signal radiative tail parameters are fixed to values obtained from simulation. The same shape parameters are used for the $\Bz\to\jpsi\pi^+\pi^-$ decays, with the mean value shifted by the known \Bs  and \Bz meson mass difference~\cite{PDG2018}. Finally, we fit simultaneously both RS and WS candidates, using the simulated shape for $\Bs\to\jpsi \eta^\prime (\to\rho^0 \gamma)$ whose yield is allowed to float, and fixing both the size and shape of the $\Lb\to\jpsi p\Km$ component. The results of the fit are shown in Fig.~\ref{fig:massfit}. We find $33\,530\pm 220$ signal \Bs within $\pm20$\mev of the $\Bs$ mass peak, with a purity of 84\%. These decays are used for further analysis. Multiple candidates in the same event have a rate of 0.20\% in a $\pm20$\mev interval around the $\Bs$ mass peak, and are retained.

To subtract the background in the signal region in the amplitude fit we add negatively weighted events from the WS sample to the RS sample, also accounting for the differing $\pi\pi$ mass and decay-time distributions. The weights are determined by comparing the RS and WS mass distributions in the upper mass sideband ($5420-5550 \mev$). In addition, a small component of \mbox{$\Bs\to\jpsi\eta^\prime (\eta^\prime \to \rho^0\gamma)$} decays is also subtracted, since it is  absent in the WS sample.

\section{Detector efficiency and resolution}

\begin{figure}[tb]
  \begin{center}
     \includegraphics[width=0.45\textwidth]{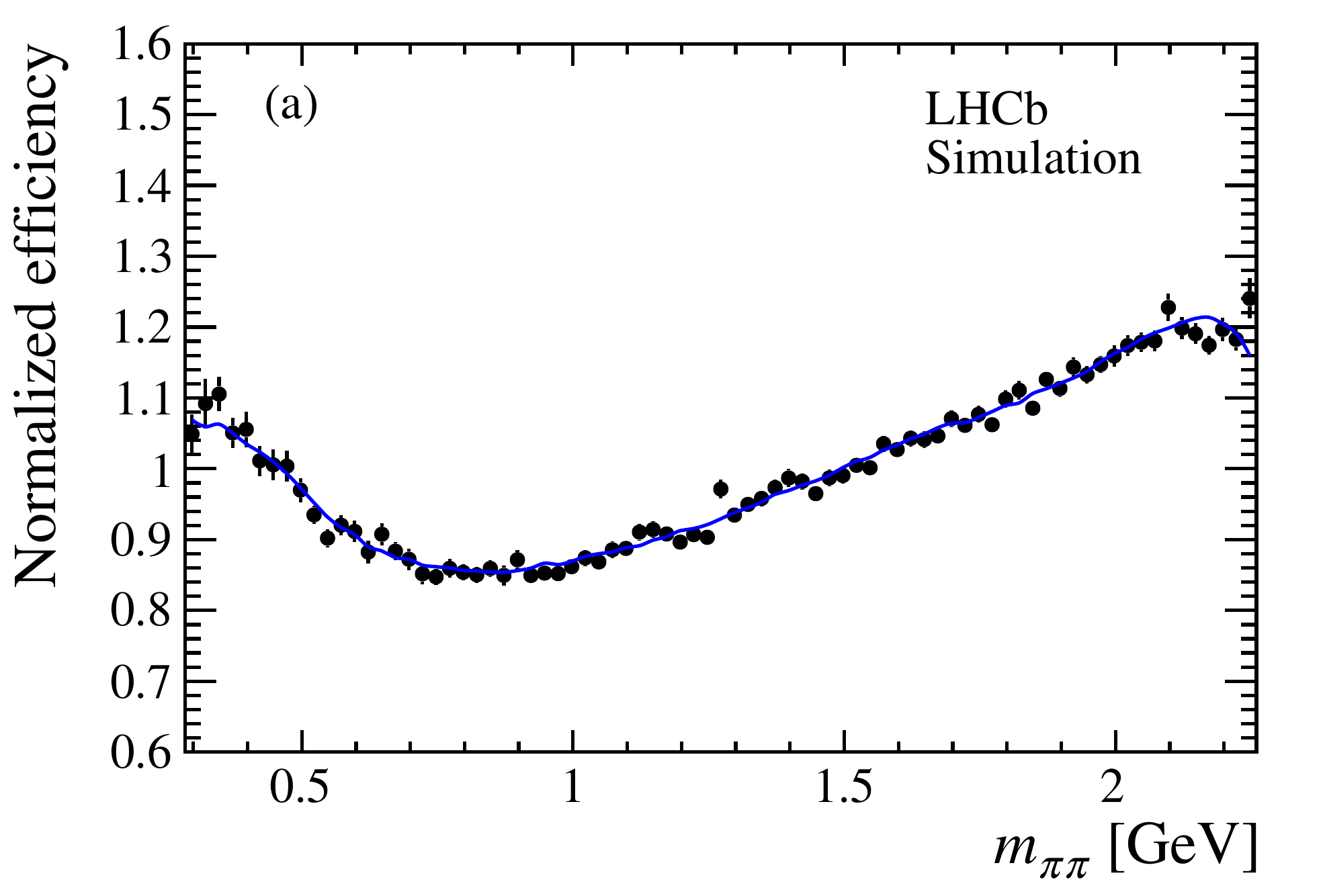}%
     \includegraphics[width=0.45\textwidth]{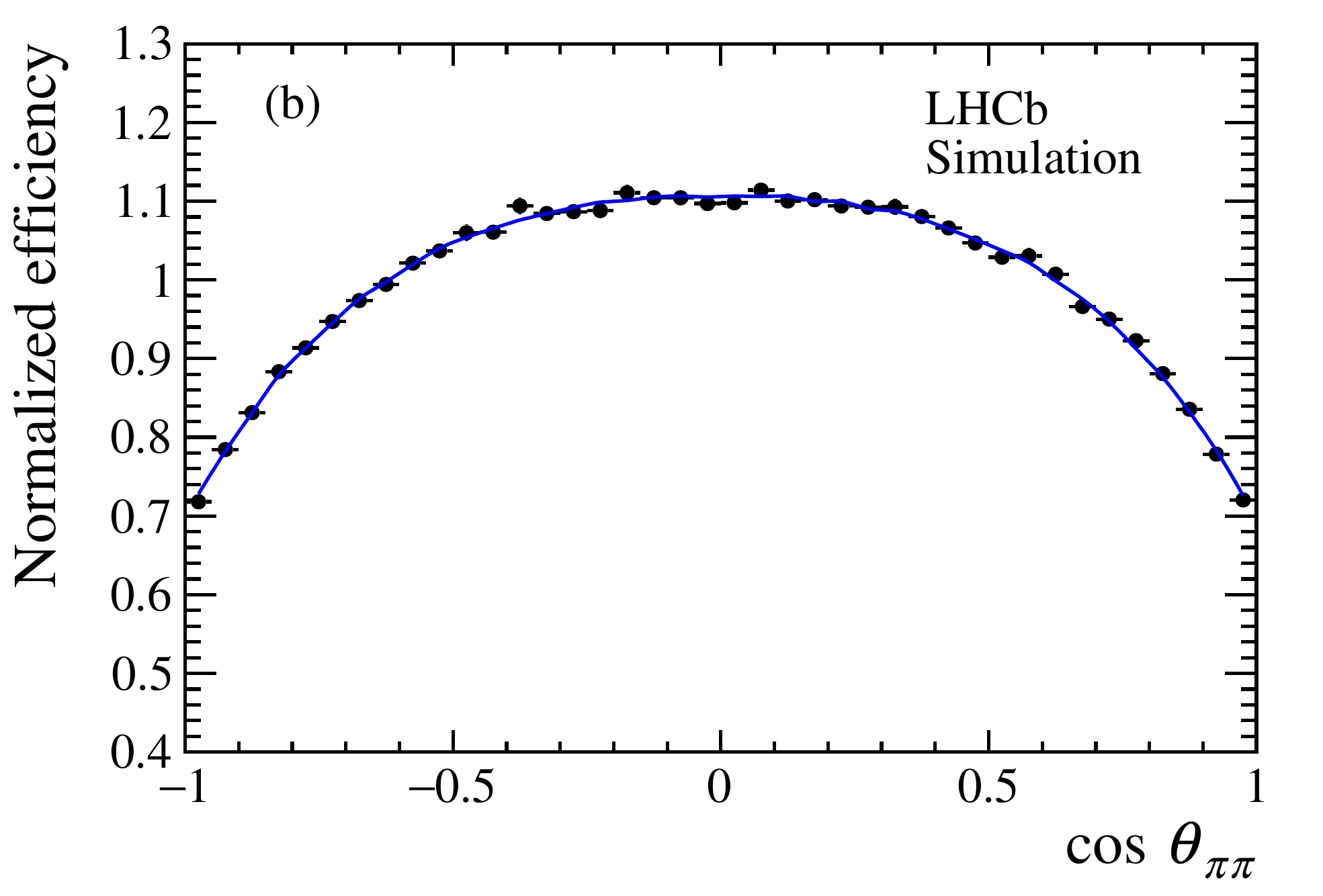}\\
      \includegraphics[width=0.45\textwidth]{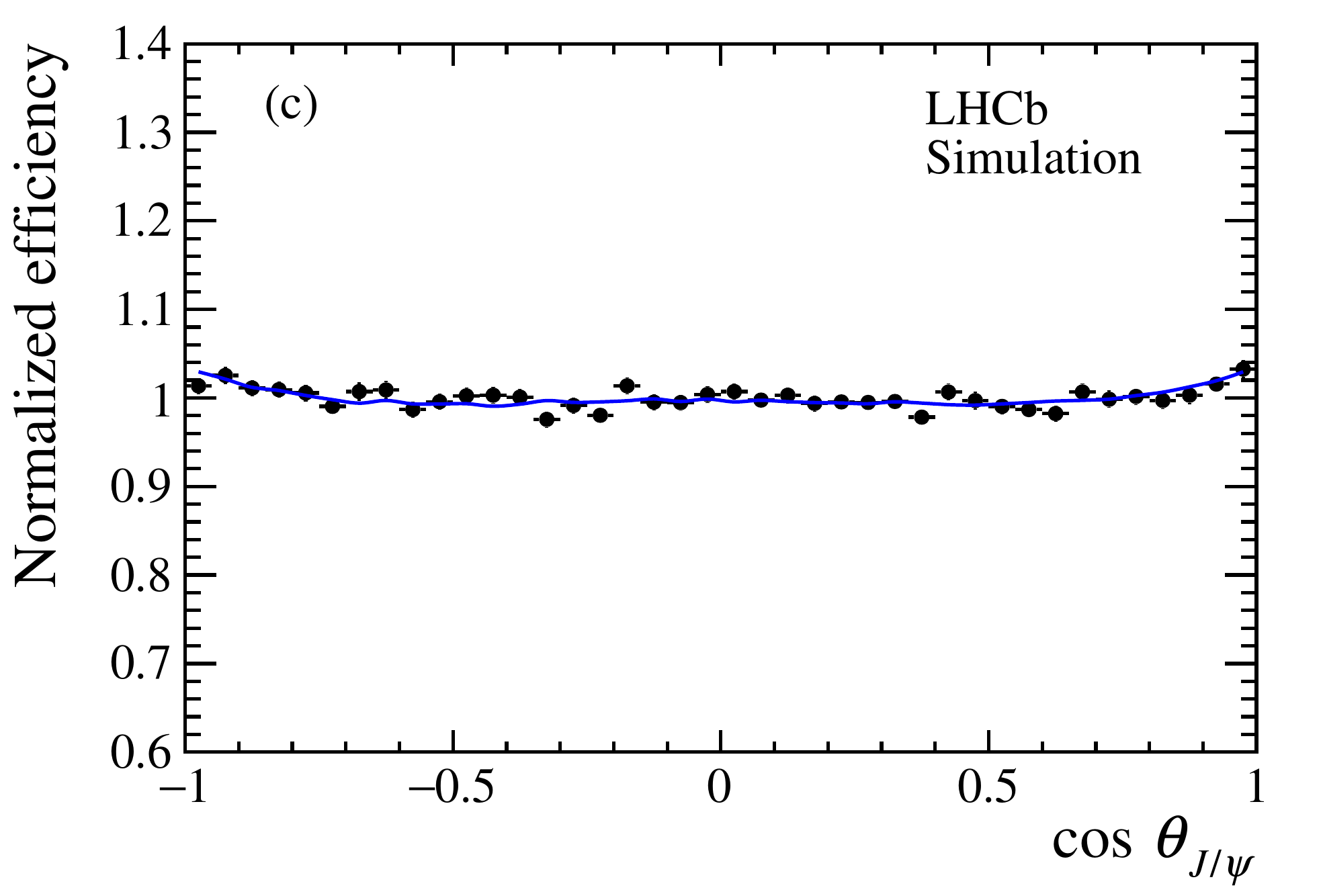}%
     \includegraphics[width=0.45\textwidth]{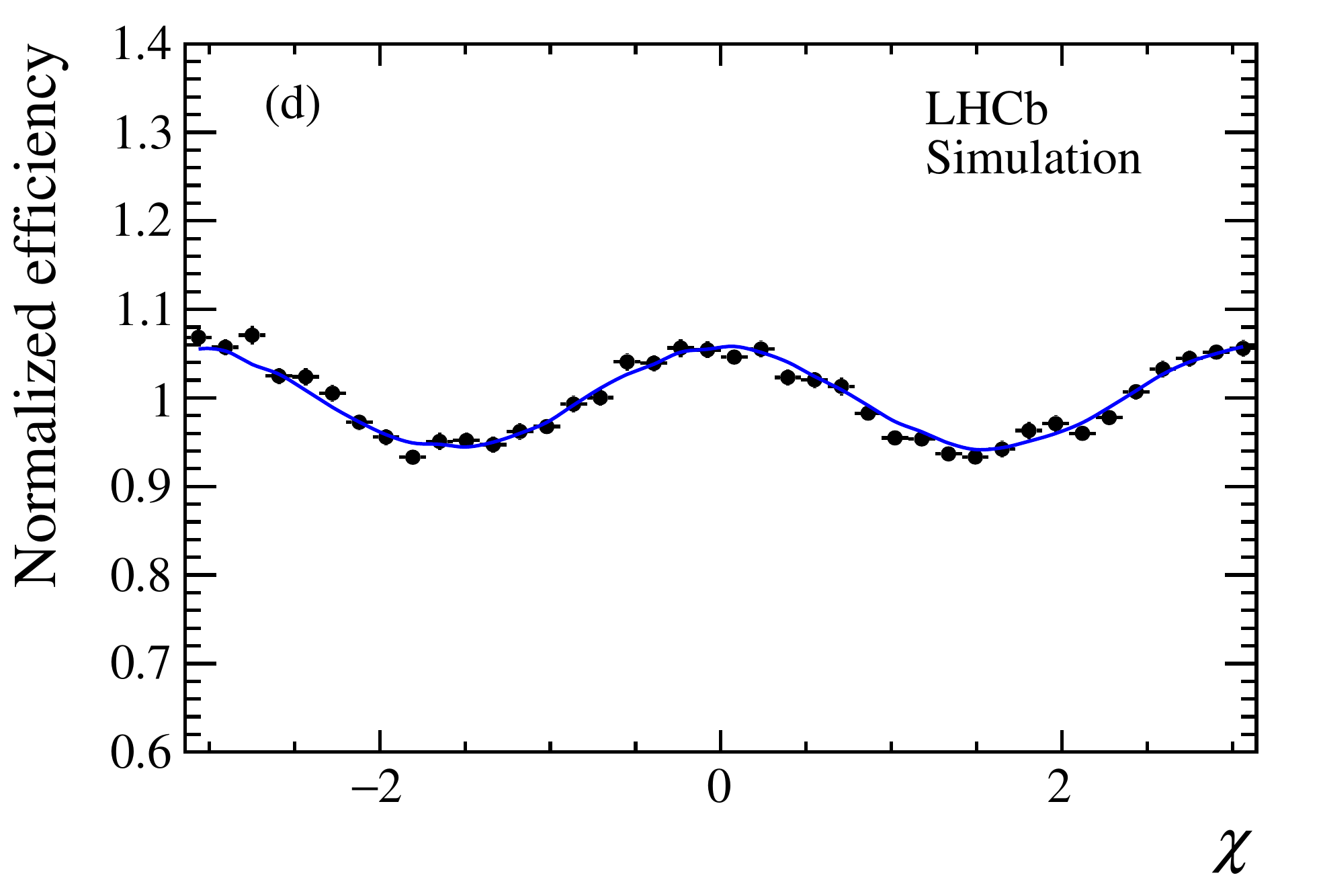}\\
    \caption{Overall efficiency normalized to unity  for (a) $\m$ , (b) $\cos \angpi$ , (c) $\cos \angmu$  and (d) $\chi$ observables. The points with error bars are from the  $\Bsjpsipipi$ simulation, while the curves show the projection of the efficiency function.}  \label{acc}
  \end{center}
\end{figure}

The correlated efficiencies  in $\m$ and angular variables $\Omega$ are determined from simulation. 
We weight the simulated signal events to reproduce the \Bs meson \pt and $\eta$ distributions as well as the track multiplicity of the events. The latter may influence the efficiencies of the tracking and particle identification. 
The calculated efficiencies are shown in Fig.~\ref{acc} along with the determined efficiency  function.
The four-dimensional efficiency is parameterized by a combination of Legendre and spherical harmonic moments \cite{Aaij:2015wza}, as
\begin{equation}\label{EqEff}
\varepsilon(\m,\cos\angpi,\cos\angmu,\chi) = \sum_{a,b,c,d}\epsilon^{abcd}P_a(\cos\angpi)Y_{bc}(\angmu,\chi)P_d\left(2\frac{\m-\m^{\rm min}}{\m^{\rm max}-\m^{\rm min}}-1\right),
\end{equation}
where $P_a$ and $P_d$ are Legendre polynomials, $Y_{bc}$ are spherical harmonics, $\m^{\rm min}=2 m_{\pi^+}$ and $\m^{\rm max}=m_{\Bs}-m_{\jpsi}$, and $\epsilon^{abcd}$ are efficiency coefficients determined from weighted averages of decays generated uniformly over phase space~\cite{LHCb-PAPER-2017-008}. 

\begin{figure}[tb]
  \begin{center}
    \includegraphics[width=0.49\textwidth]{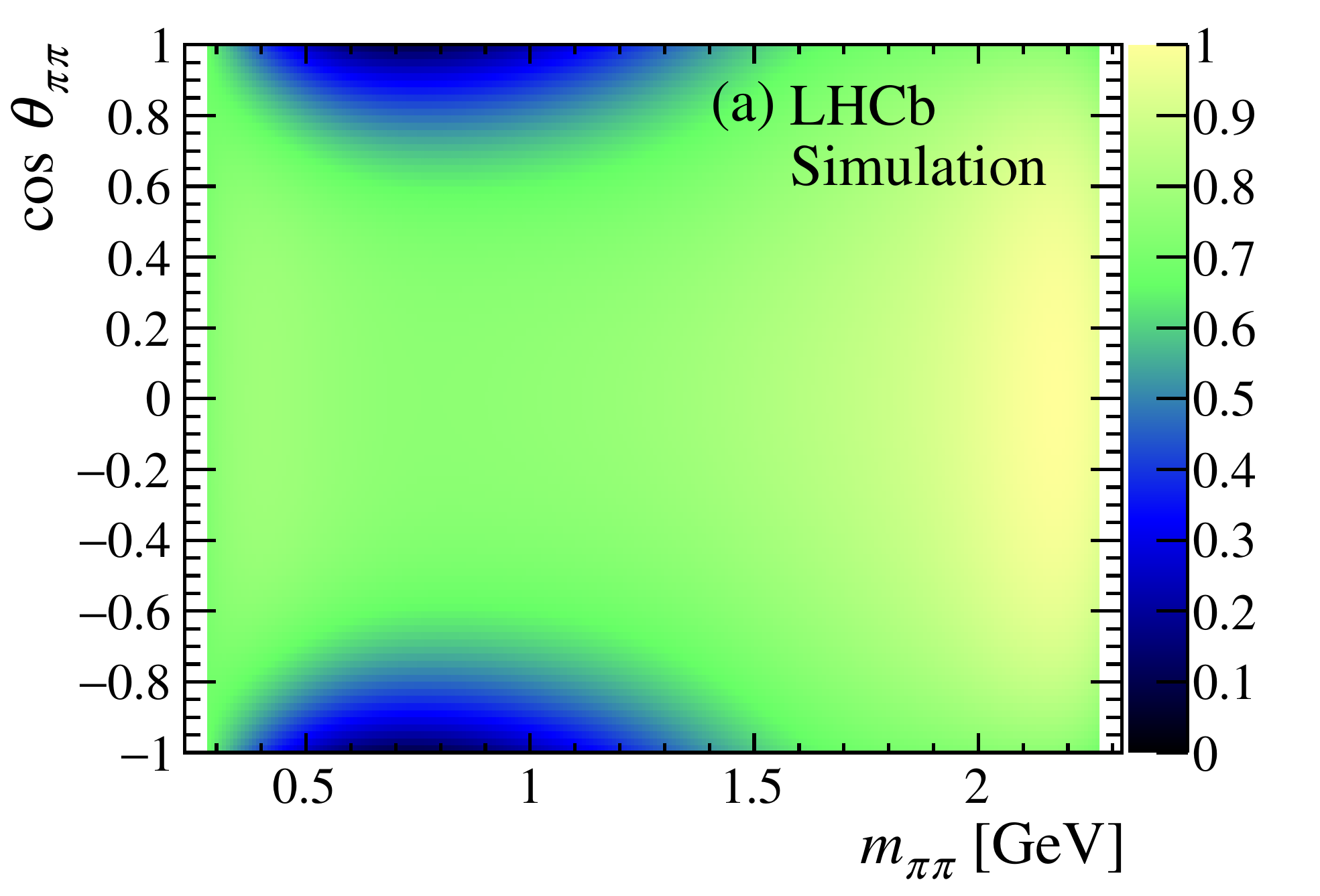}
        \includegraphics[width=0.49\textwidth]{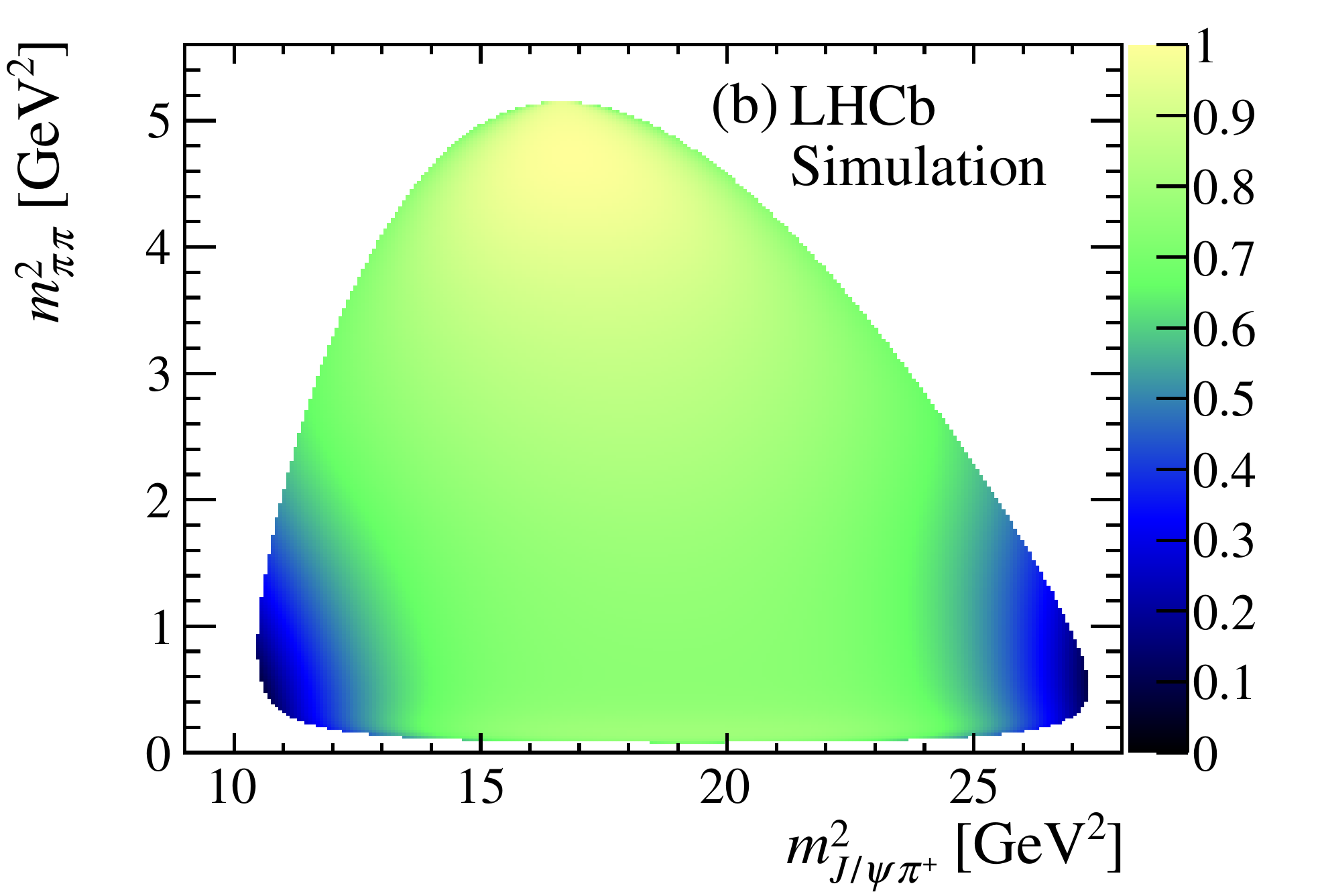}
    \caption{Overall efficiency for (a) $\m$ vs $\cos \angpi$ and (b) $m^2_{\jpsi\pip}$ vs $m^2_{\pi\pi}$. The inefficiency is at the kinematic boundary of $m^2_{\jpsi\pi^+}$ where the pion is almost at rest in the \Bs frame.}  \label{fig:eff2d}
  \end{center}
\end{figure}

The model gives an excellent representation of the simulated data.  The efficiency is uniform within about $\pm 4$\% for $\cos\angmu$ and about 10\% for $\chi$ variables; however the $\m$ and $\cos\angpi$ variables show large efficiency variations and correlations (see Fig.~\ref{fig:eff2d}), due to the \chisqip $>\ 4$ requirements on the hadrons. The loss of efficiency in the lower $\m$ region can be interpreted as the projection of the effects of cuts on \chisqip. Events at $\cos\angpi=\pm1$ and $\m\simeq0.6-0.8$\gev are at the kinematic boundary of $m^2_{\jpsi \pi^+}$. One of the pions is almost at rest in $\Bs$ rest frame, and thus the pion points to the PV, resulting in a very small \chisqip for this pion. The \chisqip variable is the most useful tool to suppress large pion combinatorial background from the PV.

The reconstruction efficiency is not constant as a function of \Bs decay time due to
displacement requirements applied to the hadrons in the offline selections and on $\jpsi$ candidates in the trigger. It is determined  using the control channel $\Bz \to \jpsi {K}^{*}(892)^0$,
with ${K}^{*}(892)^0 \to \Kp\pim$, which is known to have
a lifetime of $\tau_{\Bd} = 1.520 \pm 0.004\ps$~\cite{PDG2018}.
The simulated \Bz events are weighted to reproduce the distributions in the data for \pt and $\eta$ of the \Bz meson, and the invariant mass and helicity angle of \Kp\pim system, as well as the track multiplicity of the events. The signal efficiency is calculated as
	$\varepsilon_{\rm data}^{\Bs}(t) = \varepsilon_{\rm data}^{\Bd}(t) \cdot {\varepsilon_{\rm sim}^{\Bs}(t)}/{\varepsilon_{\rm sim}^{\Bd}(t)}$,
where $\varepsilon_{\rm data}^{\Bd}(t)$ is the efficiency of the control channel as measured by comparing data with the known lifetime distribution, and
$\varepsilon_{\rm sim}^{\Bs}(t)/\varepsilon_{\rm sim}^{\Bd}(t)$ is the ratio of efficiencies of the
simulated signal and control mode after the full trigger and selection chain have been applied.  
This correction accounts for the small differences in the kinematics between
the signal and control modes. The details of the method are explained in Ref.~\cite{LHCb-PAPER-2014-019}. 

The acceptance is checked by measuring the decay width of $\Bp\to\jpsi K^+$ decays. The fitted decay-width difference between the $\Bp$ and $\Bz$ mesons is \mbox{$\Gamma_{\Bp}-\Gamma_{\Bz} =-0.0475\pm0.0013 $\invps}, where the uncertainty is statistical only, in agreement with the known value of \mbox{$-0.0474\pm0.0023$\invps}~\cite{PDG2018}.

From the measured \Bs candidate momentum and decay distance, the decay time and its event-by-event uncertainty $\delta_t$ are calculated.  The calculated uncertainty is imbedded into the resolution function, which is modelled by the sum of three Gaussian functions with common means and widths proportional to a quadratic function of $\delta_t$. The parameters of the resolution function are determined with a sample of putative prompt $\jpsi\to\mu^+\mu^-$ decays combined with two pions of opposite charge.  Taking into account the decay-time uncertainty distribution of the \Bs signal, the average effective resolution is found to be 41.5\fs. The method is validated using simulation; we estimate the accuracy of the resolution determination to be $\pm$3\%.

\section{Flavour tagging}
Knowledge of the $\BsorBsbar$ flavour at production is necessary. We use information from decays of the other $b$ hadron in the event (opposite-side, OS) and  fragments of the jet that produced the $\BsorBsbar$ meson that contain a charged kaon, called same-side kaon (SSK) \cite{Aaij:2012mu,*Aaij:2016psi,*Fazzini:2018dyq}. The OS tagger infers the flavour of the other $b$ hadron in the event from the charges of muons, electrons, kaons, and the net charge of the particles that form reconstructed secondary vertices. 

The flavour tag, $\mathfrak{q}$, takes values of +1, $-1$ or 0 if the signal meson is tagged as $\Bs$, $\Bsb$ or untagged, respectively. The wrong-tag probability, $\mathfrak{y}$, is estimated event-by-event based on the output of a neural network. 
It is subsequently calibrated with data in order to relate it to the true wrong-tag probability of the event by a linear relation as
\begin{equation}
\begin{array}{rl}
\omega(\mathfrak{y} )&= p_0 + \frac{\Delta p_0}{2} + \left(p_1+\frac{\Delta p_1}{2}\right)\cdot(\mathfrak{y}-\langle\mathfrak{y} \rangle);\\
\overline{\omega}(\mathfrak{y}) &= p_0 - \frac{\Delta p_0}{2} + \left(p_1-\frac{\Delta p_1}{2}\right)\cdot(\mathfrak{y}-\langle\mathfrak{y} \rangle),\\
\end{array}
\end{equation}
where $p_0$,  $p_1$, $\Delta p_0$ and $\Delta p_1$ are calibration parameters, and $\omega(\mathfrak{y})$ and $\overline{\omega}(\mathfrak{y})$ are the calibrated probabilities for a wrong-tag assignment for $\Bs$ and $\Bsb$ mesons, respectively. The calibration is performed separately for the OS and the SSK taggers using $\Bp\to\jpsi \Kp$ and $\Bs\to \Dsm\pip$ decays, respectively. 
When events are tagged by both the OS and the SSK algorithms, a combined tag decision is formed. The resulting efficiency and tagging powers are listed in Table~\ref{tab:tagpower}.

\begin{table}[t]
\begin{center}
\caption{Tagging efficiency, $\varepsilon_{\rm tag}$, and tagging power given as the efficiency times dilution squared, $\varepsilon_{\rm tag}D^2$, where $D=(1-2\omega)$ for each category and the total.  The uncertainties on $\varepsilon_{\rm tag}$  are statistical only, and those for $\varepsilon_{\rm tag}D^2$ contain both statistical and systematic components.}\label{tab:tagpower}
\begin{tabular}{lcc}\hline
Category &$\varepsilon_{\rm tag}$ (\%)& $\varepsilon_{\rm tag}D^2$(\%) \\\hline
OS only &$11.0\pm0.6$&  $0.86\pm0.05$\\
SSK only & $42.6\pm0.6$&  $1.54\pm0.33$\\
OS and SSK      &  $24.9\pm0.6$& $2.66\pm0.19$ \\\hline
Total             & $78.5\pm0.7$ & $5.06\pm0.38$\\\hline
\end{tabular}
\end{center}
\end{table}

\section{\boldmath Description of the $\pi^+\pi^-$ mass spectrum}
We fit the entire $\pi^+\pi^-$ mass spectrum including the resonance contributions listed in Table~\ref{tab:res}, and a nonresonant (NR) component. We use an isobar model~\cite{LHCb-PAPER-2013-069}.  All resonances are described by Breit--Wigner amplitudes, except for the $f_0(980)$ state, which is modelled by a Flatt\'e function \cite{Flatte:1976xv}.  The nonresonant amplitude is treated as being constant in $m_{\pi\pi}$. 
Other theoretically motivated amplitude models are also proposed to described this decay~\cite{daub:2016model, ropertz:2018new}. 
The previous publication~\cite{LHCb-PAPER-2013-069}  used an unconfirmed $f_0(1790)$ resonance, reported by the BES collaboration~\cite{Ablikim:2004wn}, instead of the $f_0(1710)$ state. We test which one gives a better fit. 

\begin{table}[hbt]
\centering
\caption{Resonance parameters.}\label{tab:res}
\vspace{-0.2cm}
\begin{tabular}{cccc}
\hline
 Resonance &Mass (MeV) & Width (MeV) & Source \\
 \hline
$f_0(500)$ &$471\pm21$ &$534\pm53$ & LHCb\cite{LHCb-PAPER-2012-045}\\
$f_0(980)$ & \multicolumn{2}{c}{Varied in fits}&\\
$f_2(1270)$ & $1275.5\pm 0.8$ & $186.7_{-2.5}^{+2.2}$&PDG \cite{PDG2018}\\
$f_0(1500)$ & \multicolumn{2}{c}{Varied in fits}&\\
$f_2'(1525)$ & $1522.2\pm 1.7$ & $78.0\pm 4.8$& LHCb \cite{LHCb-PAPER-2017-008} \\
$f_0(1710)$ & $1723_{-5}^{+6}$& $139\pm8$& PDG \cite{PDG2018}\\
$f_0(1790)$ & $1790_{-30}^{+40}$ & $270_{-30}^{+60}$&BES~\cite{Ablikim:2004wn}\\
\hline
\end{tabular}\label{tab:resparam}
\end{table}

The amplitude $A_R(\m)$, generally represented by a Breit--Wigner function or a Flatt\'e function, is used to describe the mass line shape of resonance $R$. To describe the resonance from the $\Bs$ decays, the amplitude is combined with the $\Bs$ and resonance decay properties to form the following expression
\begin{equation}
\label{eq:ARmain}
{\cal A}_{R}(\m) = \sqrt{2J_R+1} \sqrt{P_R P_B}  F_B^{(L_B)} F_R^{(L_R)} A_R(\m)\left(\frac{P_B}{m_B}\right)^{L_B}\left(\frac{P_R}{m_{0}}\right)^{L_R}.
\end{equation}
Here $P_B$ is the \jpsi momentum in the $\Bs$ rest frame, $P_R$ is the momentum of either of the two hadrons in the dihadron rest frame, $m_{B}$ is the $\Bs$ mass, $m_{0}$ is the mass of resonance $R$,\footnote{Equation~(\ref{eq:ARmain}) is modified from that used in previous publications~\cite{LHCb-PAPER-2013-069,LHCb-PAPER-2014-019}  and follows the convention suggested by the PDG \cite{PDG2018}.}
 $J_R$ is the spin of the resonance $R$, $L_B$ is the orbital angular momentum between the $J/\psi$ meson and $\pip\pim$ system, and $L_R$ the orbital angular momentum in the $\pip\pim$ system, and thus is the same as the spin of the $\pip\pim$ resonance. The terms $F_B^{(L_B)}$ and $F_R^{(L_R)}$ are the Blatt--Weisskopf barrier factors for the $\Bs$ meson and $R$ resonance, respectively~\cite{LHCb:2012ae}. 
The shape parameters for the $f_0(980)$ and $f_0(1500)$ resonances are allowed to vary. 

\section{Likelihood definition}

The decay-time distribution including flavour tagging is

\begin{align}\label{eq:R}
R(\hat{t},\m,\Omega,\mathfrak{q}|\mathfrak{y}) =&\frac{1}{1+|\mathfrak{q}|}\bigg[\left[1+\mathfrak{q}\left(1-2\omega(\mathfrak{y})\right)\right]\Gamma(\hat{t},\m,\Omega)\bigg.\nonumber\\
&\bigg. +\left[1-\mathfrak{q}\left(1-2\bar{\omega}(\mathfrak{y})\right)\right]\frac{1+A_{\rm P}}{1-A_{\rm P}}\bar{\Gamma}(\hat{t},\m,\Omega)\bigg],
\end{align}
where $\hat{t}$ is the true decay time, $\brabar{\Gamma}$ is defined in Eq.~(\ref{Eq-t}), and $A_{\rm P}$ is the production asymmetry of $\Bs$ mesons.

The fit function for the signal is modified to take into account the decay-time resolution and acceptance effects resulting in
\begin{equation}
F(t,\m,\Omega,\mathfrak{q}|\mathfrak{y},\delta_t)= \left[R(\hat{t},\m,\Omega,\mathfrak{q}|\mathfrak{y}) \otimes T(t-\hat{t}|\delta_t)\right] \varepsilon_{\rm data}^{\Bs}(t) \varepsilon(\m,\Omega),
\end{equation}
where $\varepsilon(\m,\Omega)$ is the efficiency as a function of $\m$ and angular variables, $T(t-\hat{t}|\delta_t)$ is the decay-time resolution function, and $\varepsilon_{\rm data}^{\Bs}(t)$ is the decay-time acceptance function.  The free parameters in the fit are $\phi_s$, $|\lambda|$, $\GH-\Gamma_{\Bd}$, the magnitudes and phases of the resonances amplitudes, and the shape parameters of some resonances. The other parameters, including $\dms$, and \GL, are fixed to the known values \cite{PDG2018} or other measurements mentioned below.

The signal function is normalized by summing over $\mathfrak{q}$ values and integrating over decay time $t$, the mass $\m$, and the angular variables, $\Omega$, giving
\begin{equation}\label{Normsig}
{\cal N}(\delta_{t})=2\int [\Gamma(\hat{t},\m,\Omega)+\frac{1+A_{\rm P}}{1-A_{\rm P}}\bar{\Gamma}(\hat{t},\m,\Omega)] \otimes T(t-\hat{t}|\delta_{t}) \varepsilon_{\rm data}^{\Bs}(t)  \varepsilon(\m,\Omega)\,\dv \m \,\dv\Omega \,\dv t\,.
\end{equation}
We assume no asymmetries in the tagging efficiencies, which are accounted for in the systematic uncertainties.
The resulting signal PDF is 
\begin{equation}
{\cal P} (t,\m,\Omega,\mathfrak{q}|\mathfrak{y},\delta_t) = \frac{1}{{\cal N}(\delta_t)}F(t,\m,\Omega,\mathfrak{q}|\mathfrak{y},\delta_t).
\end{equation}

The fitter uses a technique similar to \sPlot \cite{Pivk:2004ty,*Xie:2009} to subtract background from the log-likelihood sum. Each candidate is assigned a weight, $W_i=+1$ for the RS events and negative values for the WS events. The likelihood function is defined as 
\begin{equation}
-2\ln {\cal L} = -2 \,s_W \sum_i W_i \ln {\cal P}(t,\m,\Omega,\mathfrak{q}|\mathfrak{y},\delta_t),
\end{equation}  
where $s_W\equiv \sum_i W_i / \sum_i W_i^2$ is a constant factor accounting for the effect of the background subtraction on the statistical uncertainty. 

The decay-time acceptance is assumed to be factorized from other variables, but due to the \chisqip cut on the two pions, the decay time is correlated with the angular variables. To avoid bias on the determination of $\GH$ from the decay-time acceptance, the simulated $\Bs$ signal is weighted in order to reproduce the $\m$ resonant structure observed in data
 by using the preferred amplitude model that is determined by the overall fit. An iterative procedure is performed to finalize the decay-time acceptance.  This procedure converges in three steps beyond which $\GH$ does not vary.  When we apply this method to pseudoexperiments that include the correlation mentioned before, the fitter reproduces the input values of $\phis$, $\GH$ and $|\lambda|$.

\section{Fit results}
We first choose the resonances that best fit the $\m$ distribution.  Table~\ref{tab:likelihood} lists the different fit components and the value of 
 $-2\ln{\cal L}$. In these comparisons, the mass and width of most resonances are fixed to the central values listed in Table~\ref{tab:res}, except for the $f_0(980)$ and $f_0(1500)$ resonances, whose parameters are allowed to vary. We find two types of fit results, one with a positive integrated sum of all interfering components and one with a negative one. 
 The first listed Solution I is better than Solution II by four standard deviations, calculated by taking the square root of the $-2\ln{\cal L}$ difference. We take Solution I for our measurement and II for systematic uncertainty evaluation. The models corresponding to Solutions I and II are very similar to those found in our previous analysis of the same final state~\cite{LHCb-PAPER-2013-069}.
\begin{table}[t]
\centering
\caption{Likelihoods of various resonance model fits. Positive or negative interferences (Int) among the contributing resonances are indicated. The Solutions are indicated by \#.}\label{tab:likelihood}
\begin{tabular}{rlcl}
\hline
\# & Resonance content& Int&$-2\ln{\cal L}$ \\\hline
I&$f_0(980)+f_0(1500)+f_0(1790)+f_2(1270)+f_2^\prime(1525)$+NR &$-$& {$-4850$} \\
II&$f_0(980)+f_0(1500)+f_0(1710)+f_2(1270)+f_2^\prime(1525)+$NR &+&  {$-4834$} \\
III&$f_0(980)+f_0(1500)+f_0(1790)+f_2(1270)+f_2^\prime(1525)+$NR &+&{$-4830$} \\
IV&$f_0(980)+f_0(1500)+f_0(1790)+f_2(1270)+f_2^\prime(1525)$ &$-$& {$-4828$} \\
V&$f_0(980)+f_0(1500)+f_0(1710)+f_2(1270)+f_2^\prime(1525)$ & $-$&$-4706$ \\\hline
\end{tabular}
\end{table}

For the fit we assume that the \CP-violation quantities (${\phi_s}_i$, $|\lambda_i|$) are the same for all the resonances. 
We also fix $\dms$ to the central value of the world average \mbox{$17.757\pm 0.021$\invps}~\cite{PDG2018},  and fix  $\GL$ to the central value of $0.6995\pm0.0047$\invps from the LHCb $\Bs\to \jpsi K^+K^-$ results~\cite{LHCb-PAPER-2017-008}. 

\begin{figure}[tb]
\centering
\includegraphics[width=0.5\textwidth]{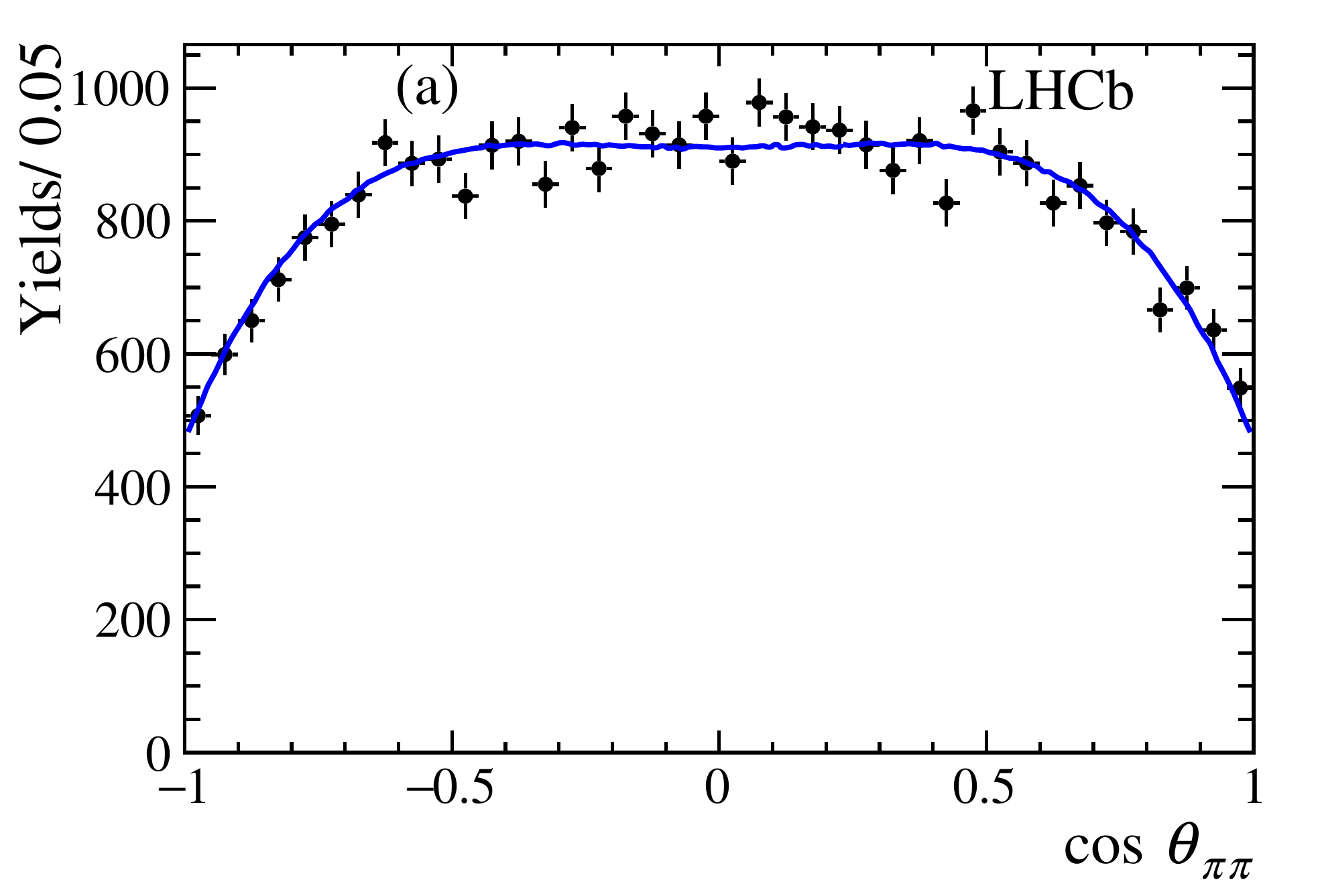}%
\includegraphics[width=0.5\textwidth]{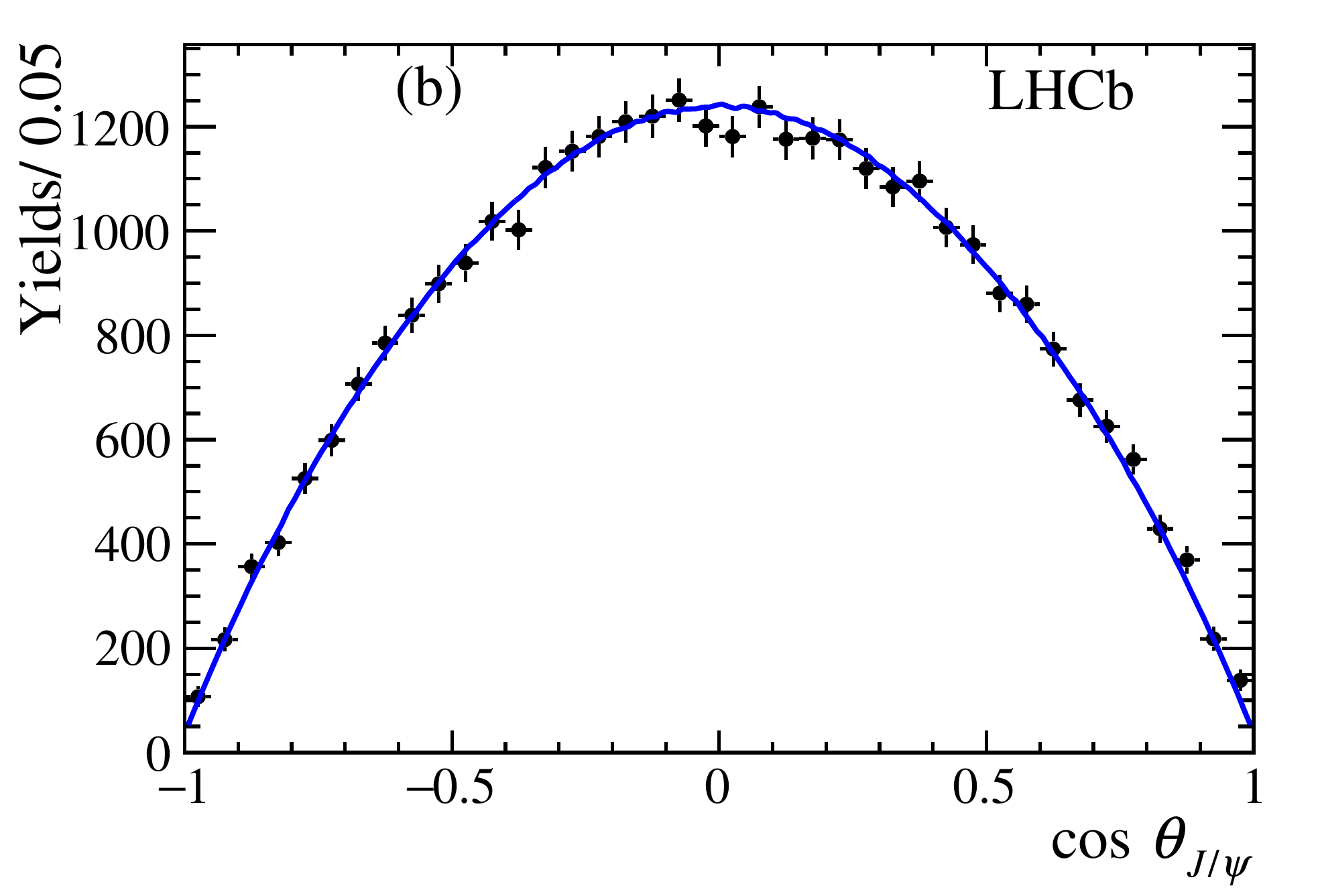}
\includegraphics[width=0.5\textwidth]{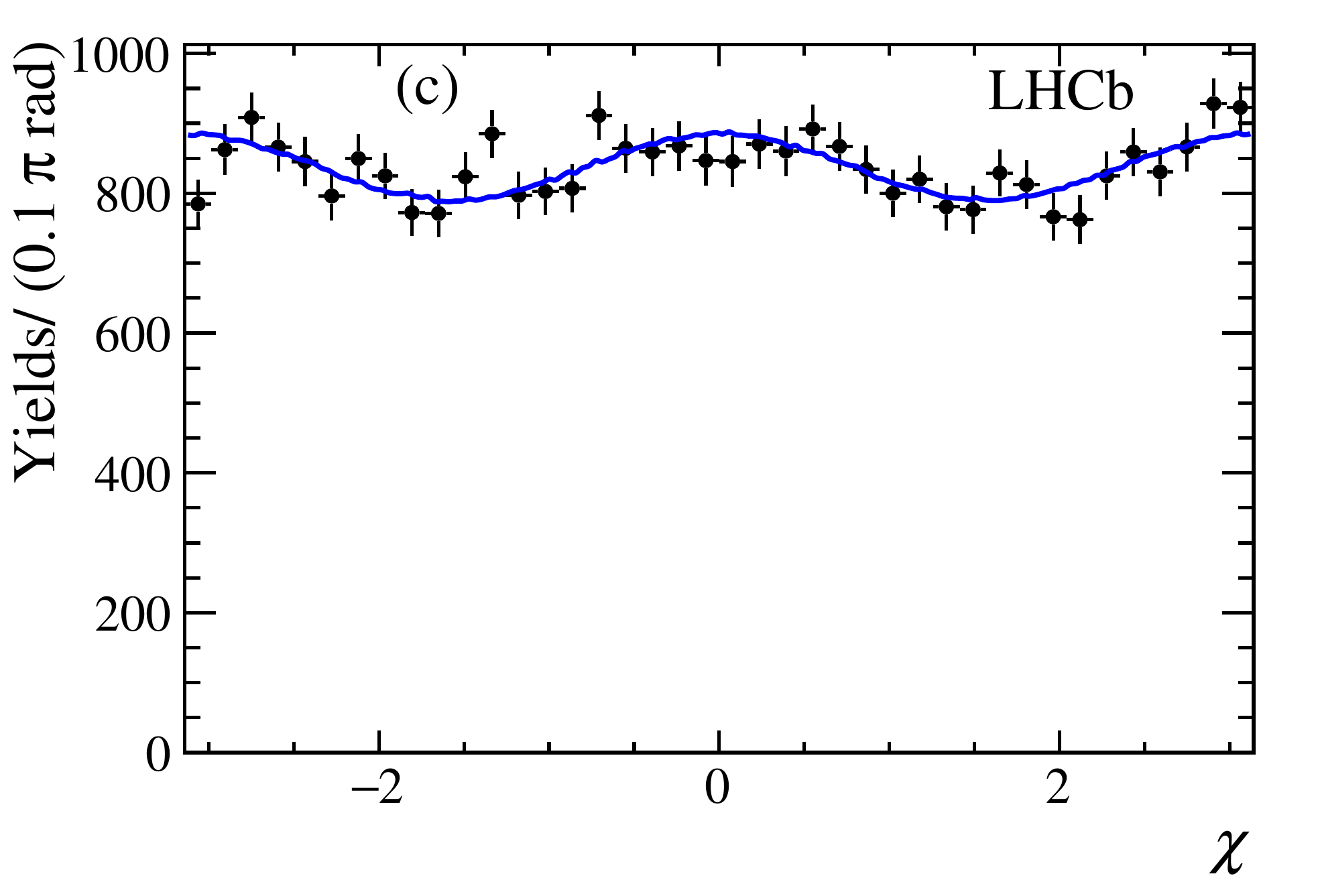}%
\includegraphics[width=0.5\textwidth]{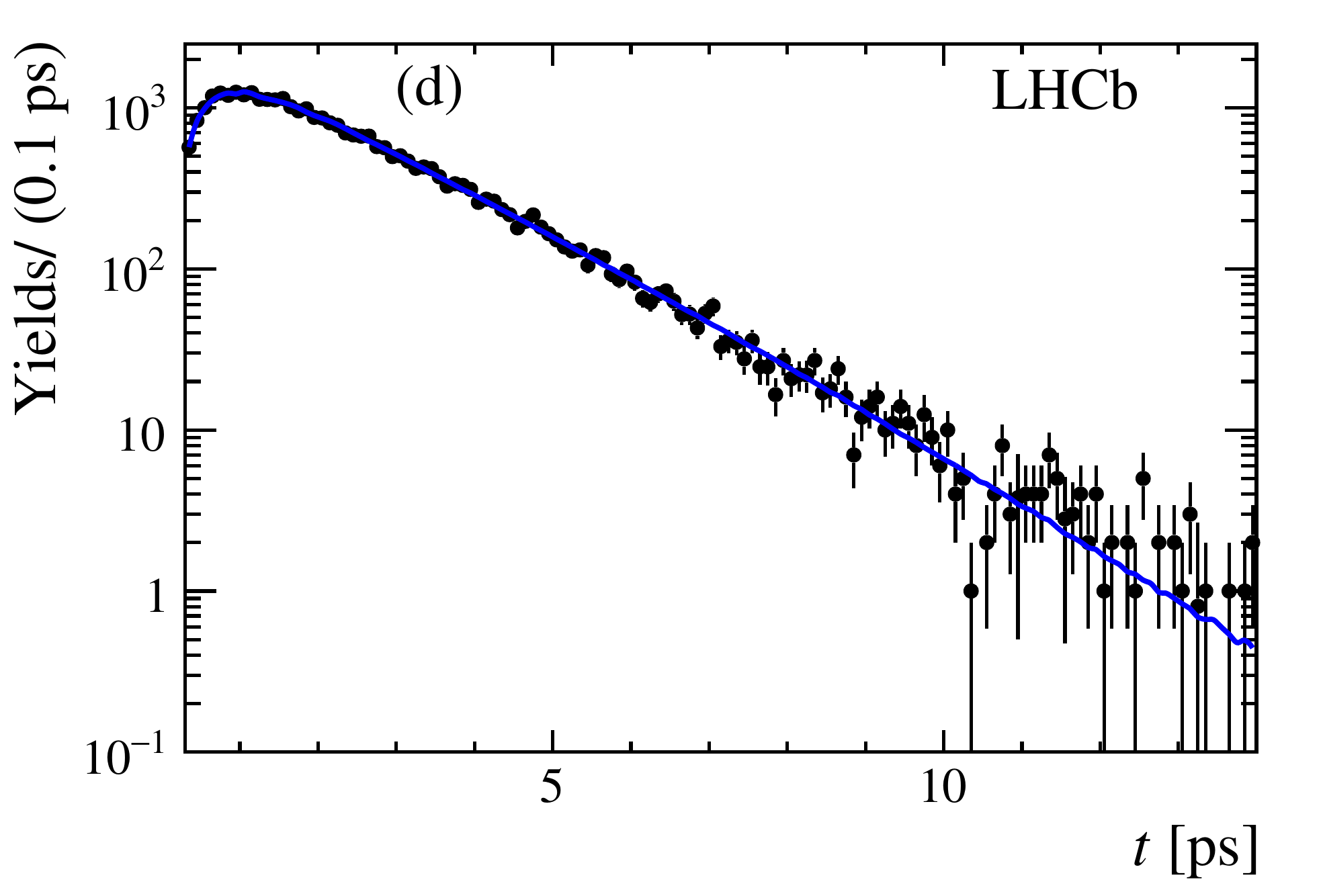}
\caption{Projections of the angular and decay-time variables with the fit result overlaid. The points (black) show the data and the curves (blue) the fits. }
\label{fig:fithigh}
\end{figure}

\begin{figure}[b]
\centering
\includegraphics[width=0.7\textwidth]{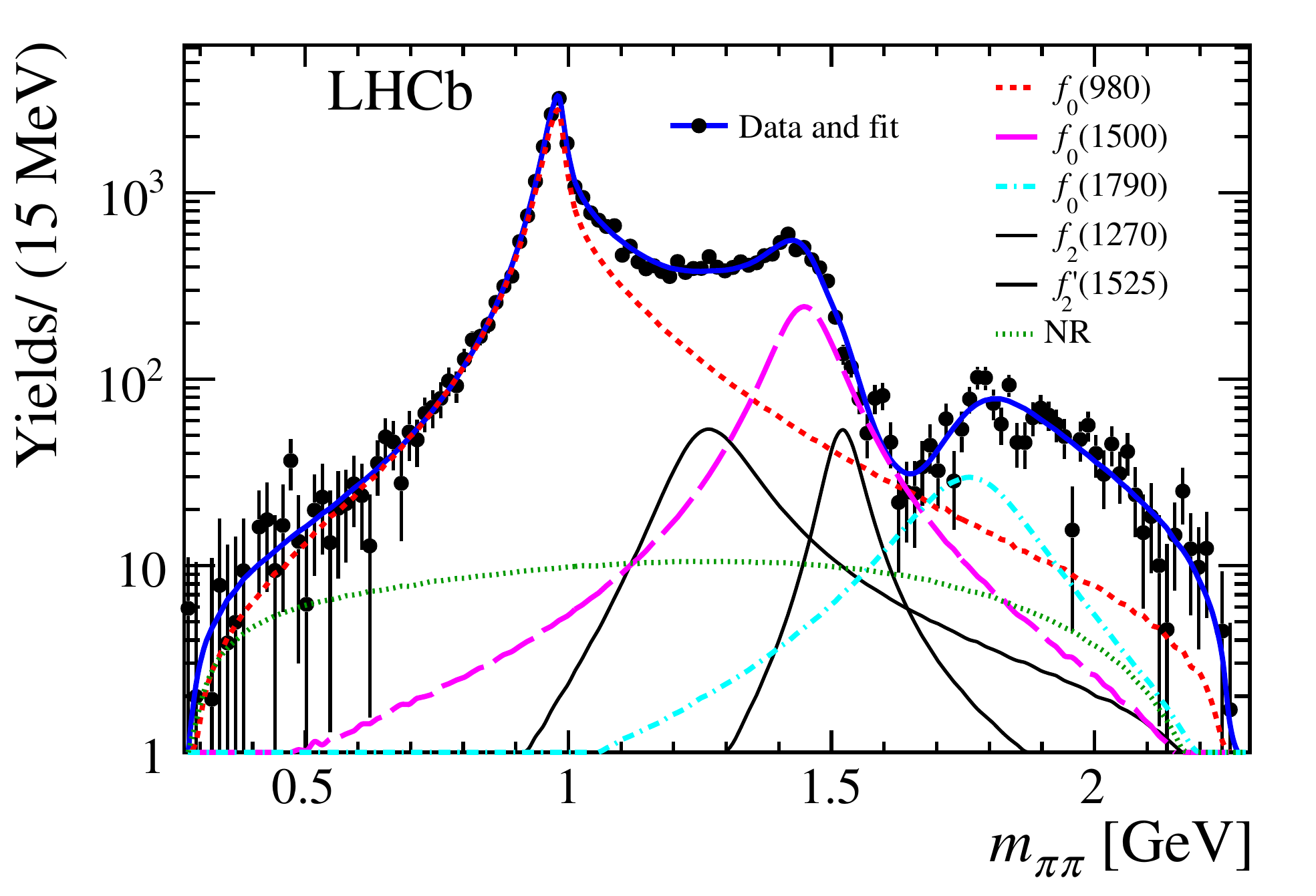}%
\caption{Data distribution of $\m$ with the projection of the Solution I fit result overlaid. The data are described by the points (black) with error bars. The solid (blue) curve shows the overall fit. }
\label{Fig0_New}
\end{figure}

The fit values and correlations of the \CP-violating parameters are shown in Table~\ref{tab:fitub} for Solution I. 
The shape parameters of $f_0(980)$ and $f_0(1500)$ resonances are found to be consistent with our previous results~\cite{LHCb-PAPER-2013-069}. 
The angular and decay-time fit projections are shown in Fig.~\ref{fig:fithigh}. The $m_{\pi\pi}$ fit projection is shown in Fig.~\ref{Fig0_New}, where the contributions of the individual resonances are also displayed. All solutions listed in Table~\ref{tab:likelihood} give very similar fit values for $\phis$ and $\GH$. We also find that the \CP-odd fraction is greater than $97$\% at 95\% confidence level. 
The resonant content for Solutions I and II are listed in Table~\ref{tab:fit2}.

\begin{table}[b]
\centering
\caption{Fit results for the \CP-violating parameters for Solution I. The first uncertainties are statistical, and the second systematic. The last three columns show the statistical correlation coefficients for the three parameters.}\label{tab:fitub}
\begin{tabular}{lcccc}
\hline
& Fit result& \multicolumn{3}{c}{Correlation}\\\hline
Parameter & &$\GH-\Gamma_{\Bz}$& $|\lambda|$ & $\phi_s$ \\\hline
$\GH-\Gamma_{\Bz} $ (\invps)& $-0.050\pm0.004\pm0.004$ & 1.000 & 0.022 & 0.038\\
$|\lambda|$& $1.01_{-0.06}^{+0.08}\pm0.03$ & 0.022 & 1.000 & 0.065 \\
$\phi_s$ (rad)& $-0.057\pm0.060\pm0.011$& 0.038 & 0.065 & 1.000\\\hline
\end{tabular}
\end{table}

\begin{table}[bt]
\centering
\caption{Fit results of the resonant structure for both Solutions I and II. These results do not supersede those in Ref.~\cite{LHCb-PAPER-2013-069} for the resonant fractions because no systematic uncertainties are quoted. The sum of fit fraction is not necessary 100\% due to possible interferences between resonances with the same spin.}\label{tab:fit2}
\def\arraystretch{1.2}
\begin{tabular}{l
cccc}
\hline
{Component}& {Fit fractions (\%)} &\multicolumn{3}{c}{Transversity fractions (\%)}\\
&  &
                    $0$ & $\parallel$& $\perp$\\\hline
\multicolumn{5}{c}{Solution I}\\\hline
$f_0(980)$	&	$\!\!\!60.09	\pm	1.48	$&$	100 			$&$	-			$&$	-			$\\
$f_0(1500)$	&	$8.88	\pm	0.87	$&$	100 			$&$	-			$&$	-			$\\
$f_0(1790)$	&	$1.72	\pm	0.29	$&$	100 			$&$	-			$&$	-			$\\
$f_2(1270)$	&	$3.24	\pm	0.48	$&$	\!\!\!13 	\pm	3 	$&$	\!\!\!37 	\pm	9 	$&$	50 	\pm	10 	$\\
$f_2^\prime(1525)$	&$	1.23	\pm	0.86	$&$	40 	\pm	13 	$&$	31 	\pm	14 	$&$	29 	\pm	25 	$\\
NR	&$	2.64	\pm	0.73	$&$	100 			$&$	-			$&$	-			$\\
\hline
\multicolumn{5}{c}{Solution II}\\\hline
$f_0(980)$	&$	\!\!\!93.05 	\pm	1.12 	$&$	100 			$&$	-			$&$	-			$\\
$f_0(1500)$	&$	6.47 	\pm	0.41 	$&$	100 			$&$	-			$&$	-			$\\
$f_0(1710)$	&$	0.74 	\pm	0.11 	$&$	100 			$&$	-			$&$	-			$\\
$f_2(1270)$	&$	3.22 	\pm	0.44 	$&$	17 	\pm	4 	$&$	\!\!\!30 	\pm	8 	$&$	53 	\pm	10 	$\\
$f_2^\prime(1525)$	&$	1.44 	\pm	0.36 $	&$	35 	\pm	8 	$&$	31 	\pm	12 	$&$	34 	\pm	17 	$\\
NR	&$	8.13 	\pm	0.79 	$&$	100 			$&$	-			$&$	-			$\\\hline
\end{tabular}
 \end{table}
  
\section{Systematic uncertainties}
\label{sec:sys}

The systematic uncertainties for the \CP-violating parameters, $\lambda$ and $\phi_s$, are smaller than the statistical ones. They are summarized in Table~\ref{tab:sys1} along with the uncertainty on $\GH-\Gamma_{\Bd}$.
The uncertainty on the decay-time acceptance is found by varying the parameters of the acceptance function within their uncertainties and repeating the fit. The same procedure is followed for the uncertainties on the \Bz lifetime,  $\Delta m_s$, $\Gamma_L$, $\m $ and angular efficiencies, resonance masses and widths, flavour-tagging calibration, and allowing for a 2\% production asymmetry~\cite{LHCb-PAPER-2016-062}; this uncertainty also includes any possible difference in flavour tagging between \Bsb and \Bs.
Simulation is used to validate the method for the time-resolution calibration. The uncertainties of the parameters of the time-resolution model are estimated using the difference between the signal simulation and prompt $\jpsi$ simulation. These uncertainties are varied to obtain the effects on the physics parameters. 
Resonance modelling uncertainty includes varying the Breit--Wigner barrier factors, changing the default values of $L_B=1$ for the D-wave resonances to one or two, the differences between the two best solutions,  and replacing the NR component by the $f_0(500)$ resonance. Furthermore,  including an isospin-violating $\rho(770)^0$ component in the fit, results in a negligible contribution of  $(1.1\pm0.3)$\%.  The largest shift among the modelling variations is taken as systematic uncertainty. The inclusion of $\rho$ components results in the largest shifts of the three physics parameters quoted. The process $B_c^+\to \pi^+\Bs$  can affect the measurement of $\GH-\Gamma_{\Bz}$. An estimate of the fraction of these decays in our sample is 0.8\%~\cite{LHCb-PAPER-2014-059}. 
Neglecting the $\Bc$ contribution leads to a bias of 0.0005\invps, which is added as a systematic uncertainty. Other parameters are unchanged. 

Corrections from penguin amplitudes are ignored because their effects are known to be small \cite{Fleischer:2015mla,LHCb-PAPER-2014-058,LHCB-PAPER-2015-034} compared to the current experimental precision. 

\begin{table}[t]
\centering
\caption{\small Absolute systematic uncertainties for the physics parameters. }\label{tab:sys1}
\begin{tabular}{lccc}
\hline
Source &	$\GH-\Gamma_{\Bz}$&	$|\lambda|$ &	$\phis$\\
 &  [{\ensuremath{\fs^{-1}}\xspace}] & $[\times 10^{-3}]$ & [mrad] \\\hline
Decay-time acceptance &$	2.0	$&$	0.0	$&$	0.3	$	\\	
$\tau_\Bz$	&$	0.2	$&$	0.5	$&$	0.0	$	\\	
Efficiency	($\m$, $\Omega$)&$	0.2	$&$	0.1	$&$	0.0	$	\\	
Decay-time resolution width&$	0.0	$&$	4.3	$&$	4.0	$	\\
Decay-time resolution mean&$	0.3	$&$	1.2	$&$	0.3	$	\\
Background	&$	3.0	$&$	2.7	$&$	0.6	$	\\
Flavour tagging	&$	0.0	$&$	2.2	$&$	2.3	$\\
$\dms$& $	0.3	$&$	4.6	$&$	2.5	$\\
$\GL$&$	0.3	$&$	0.4	$&$	0.4	$\\
$\Bc$      &  $0.5$ & - &  - \\
Resonance parameters &  0.6 	& 1.9 & 	0.8\\ 
Resonance modelling	&$	0.5	$&$	28.9$&$	9.0	$	\\
Production asymmetry& 0.3 & 0.6 & 3.4 \\
\hline
Total&$	3.8	$&$	29.9	$&$	11.0$	\\
\hline

\end{tabular}
\end{table}

\section{Conclusions}
Using $\Bs$ and $\Bsb\to\jpsi\pip\pim$ decays, we measure the \CP-violating phase, \mbox{$\phis =-0.057\pm0.060\pm0.011$~rad}, the decay-width difference 
\mbox{$\GH-\Gamma_{\Bz} =-0.050\pm0.004\pm0.004$~\!\invps}, and the parameter \mbox{$|\lambda|=1.01_{-0.06}^{+0.08}\pm0.03$}, where the quoted uncertainties are  statistical and systematic. These results are more precise than those obtained from the previous study of this mode using 7\tev and 8\tev $pp$ collisions (Run 1) \cite{LHCb-PAPER-2014-019}. 
To combine the Run-1 results with these, we reanalyze them by fixing $\dms= 17.757\pm 0.021$\invps from Ref.~\cite{PDG2018}, and \mbox{$\GL=0.6995\pm0.0047$\invps} from the LHCb $\Bs\to \jpsi K^+K^-$ results~\cite{LHCb-PAPER-2017-008}. We remove the Gaussian constraint on $\DGs$ and let $\GH$ vary. Instead of taking the uncertainties of flavour tagging and decay-time resolution into the statistical uncertainty, we place these sources in the systematic uncertainty and assume 100\% correlation with our new results. The updated results are: $\phis=0.075\pm0.065\pm0.014$\,rad and $|\lambda|=0.898\pm0.051\pm0.013$ with a correlation of $0.025$. We then use the updated $\phis$ and $|\lambda|$ Run-1 results as a constraint into our new $\phis$ fit.\footnote{We do not include an average value of $\GH$ since no systematic uncertainty was assigned for the Run-1 result.} The combined results  are $\GH-\Gamma_{\Bz}=-0.050\pm0.004\pm0.004$~\!\invps, $|\lambda| =0.949\pm0.036\pm0.019$, and $\phi_s =0.002\pm0.044\pm0.012$~\!rad. The correlation coefficients among the fit parameters are 0.025 $(\rho_{12})$, --0.001 $(\rho_{13})$, and 0.026 $(\rho_{23})$.

Our results still have uncertainties greater than the SM prediction and are slightly more precise than the measurement using $\Bs\to\jpsi K^+K^-$ decays, based only on Run-1 data, which has a precision of 0.049~rad~\cite{LHCb-PAPER-2014-059}. Hence this is the most precise determination of $\phi_s$ to date.

\section*{Acknowledgements}
\noindent We express our gratitude to our colleagues in the CERN
accelerator departments for the excellent performance of the LHC. We
thank the technical and administrative staff at the LHCb
institutes.
We acknowledge support from CERN and from the national agencies:
CAPES, CNPq, FAPERJ and FINEP (Brazil);
MOST and NSFC (China);
CNRS/IN2P3 (France);
BMBF, DFG and MPG (Germany);
INFN (Italy);
NWO (Netherlands);
MNiSW and NCN (Poland);
MEN/IFA (Romania);
MSHE (Russia);
MinECo (Spain);
SNSF and SER (Switzerland);
NASU (Ukraine);
STFC (United Kingdom);
NSF (USA).
We acknowledge the computing resources that are provided by CERN, IN2P3
(France), KIT and DESY (Germany), INFN (Italy), SURF (Netherlands),
PIC (Spain), GridPP (United Kingdom), RRCKI and Yandex
LLC (Russia), CSCS (Switzerland), IFIN-HH (Romania), CBPF (Brazil),
PL-GRID (Poland) and OSC (USA).
We are indebted to the communities behind the multiple open-source
software packages on which we depend.
Individual groups or members have received support from
AvH Foundation (Germany);
EPLANET, Marie Sk\l{}odowska-Curie Actions and ERC (European Union);
ANR, Labex P2IO and OCEVU, and R\'{e}gion Auvergne-Rh\^{o}ne-Alpes (France);
Key Research Program of Frontier Sciences of CAS, CAS PIFI, and the Thousand Talents Program (China);
RFBR, RSF and Yandex LLC (Russia);
GVA, XuntaGal and GENCAT (Spain);
the Royal Society
and the Leverhulme Trust (United Kingdom);
Laboratory Directed Research and Development program of LANL (USA).

\addcontentsline{toc}{section}{References}
\bibliographystyle{LHCb}
\bibliography{standard,LHCb-PAPER,LHCb-CONF,LHCb-DP,LHCb-TDR,CP}
\newpage
\centerline
{\large\bf LHCb Collaboration}
\begin
{flushleft}
\small
R.~Aaij$^{29}$,
C.~Abell{\'a}n~Beteta$^{46}$,
B.~Adeva$^{43}$,
M.~Adinolfi$^{50}$,
C.A.~Aidala$^{77}$,
Z.~Ajaltouni$^{7}$,
S.~Akar$^{61}$,
P.~Albicocco$^{20}$,
J.~Albrecht$^{12}$,
F.~Alessio$^{44}$
% ...

{\footnotesize \it

% ...
$ ^{7}$Universit{\'e} Clermont Auvergne, CNRS/IN2P3, LPC, Clermont-Ferrand, France\\
% ...
$ ^{12}$Fakult{\"a}t Physik, Technische Universit{\"a}t Dortmund, Dortmund, Germany\\
% ...
$ ^{20}$INFN Laboratori Nazionali di Frascati, Frascati, Italy\\
% ...
$ ^{29}$Nikhef National Institute for Subatomic Physics, Amsterdam, Netherlands\\
% ...
$ ^{43}$Instituto Galego de F{\'\i}sica de Altas Enerx{\'\i}as (IGFAE), Universidade de Santiago de Compostela, Santiago de Compostela, Spain\\
$ ^{44}$European Organization for Nuclear Research (CERN), Geneva, Switzerland\\
% ...
$ ^{46}$Physik-Institut, Universit{\"a}t Z{\"u}rich, Z{\"u}rich, Switzerland\\
% ...
$ ^{50}$H.H. Wills Physics Laboratory, University of Bristol, Bristol, United Kingdom\\
% ...
$ ^{61}$University of Cincinnati, Cincinnati, OH, United States\\
% ...
$ ^{63}$Syracuse University, Syracuse, NY, United States\\
% ...
$ ^{77}$University of Michigan, Ann Arbor, United States, associated to $^{63}$\\
% ...
}
\end{flushleft}

\end{document}